\documentclass[11pt, reqno]{amsart} 

\usepackage[]{graphicx}
\usepackage[]{xcolor}
\usepackage{alltt}
\usepackage[margin=1in]{geometry}
\usepackage{amsmath}
\usepackage{natbib}
\usepackage[title]{appendix}
\usepackage{bm}
\usepackage{amsfonts}
\usepackage[hidelinks,urlcolor=magenta]{hyperref}
\usepackage{multirow}
\usepackage{amsaddr}
\usepackage{bbm}
\usepackage{ulem} 
\usepackage{cancel} 
\usepackage[mathlines]{lineno}

\title[Dynamic spatial-temporal ordinal models]{Dynamic Spatio-Temporal Sequential Ordinal Models: Application to Invasive Weeds}

\author[Hosack, Yang, Pike, O'Loughlin, Cook, Howland, Sutcliffe, Milner, Gooden, Froese]{Geoffrey R. Hosack${}^1$, Wen-Hsi Yang${}^2$, Kyana N. Pike${}^3$, Luke S. O'Loughlin${}^4$, Emma Cook${}^4$, Brett Howland${}^4$, Emily Sutcliffe${}^5$, Richard N. C. Milner${}^5$, Ben Gooden${}^6$ \and Jens G. Froese${}^7$}
\address{${}^1$CSIRO Data61, Hobart, Tasmania,  Australia\\
${}^2$CSIRO Data61,  Dutton Park, Queensland,  Australia\\
${}^3$CSIRO Health and Biosecurity, Townsville, Queensland,  Australia\\
${}^4$Office of Nature Conservation, Australian Capital Territory Government, Canberra, Australian Capital Territory, Australia\\
${}^5$Parks and Conservation Service, Australian Capital Territory Government, Canberra, Australian Capital Territory,  Australia\\
${}^6$CSIRO Health and Biosecurity, Canberra,  Australian Capital Territory, Australia\\
${}^7$CSIRO Health and Biosecurity,  Dutton Park,  Queensland, Australia
}
\thanks{Corresponding author: Geoffrey R. Hosack. Email address: Geoff.Hosack@csiro.au}

\begin{document}


\begin{abstract}
	
	The multivariate sequential ordinal model is investigated for use in the Bayesian analysis of spatio-temporal ordinal data. The sequential ordinal model likelihood is equivalent to a binary model conditional on unknown regression coefficients and spatio-temporal random effects. Therefore, estimation and prediction in the space-time context can proceed using the well-established dynamic generalised linear model framework. Moreover, the sequential ordinal model avoids the ordering constraints on the threshold parameters that determine the category break points required by cumulative ordinal models, and so simplifies the estimation procedure for high-dimensional space-time applications using Bayesian inference. The dynamic spatio-temporal sequential ordinal model is applied to estimate foliage cover abundance of four actively managed invasive alien species. These invasive weed species are observed by means of a modified Braun-Blanquet score that is commonly used in vegetation studies and constitutes ordinal data. The multivariate ordinal data for the managed weeds species are sparsely distributed in space and time with few observations recorded in high foliage cover categories. A separable model for space-time dependence that maintains parameter interpretability in the presence of aggregated ordinal categories is therefore developed. Estimation and prediction is demonstrated using integrated nested Laplace approximation (INLA) methods developed for univariate spatio-temporal models. Bayesian estimation and prediction shows that the four invasive weed species differentially respond to habitat type, control effort and accessibility, and share similar magnitudes of dependence with short effective spatial ranges and strong temporal autocorrelations.
	
	\smallskip
	\noindent \textbf{Keywords:} collapsibility,  Gompertz population growth, grouped Cox model, proportional hazards model, species distribution model
\end{abstract}

\maketitle

\section{Introduction}

There is an increasing need for spatio-temporal prediction and inference of ordinal data. Ordinal data commonly arise in expert assessments, for example, the Likert scale for measuring attitude \citep{Gob2007}, and the applications for guiding decisions using such data are manifold. Spatial studies of ordinal data are found in diverse domains such as vegetation ecology \citep{Guisan2000}, environmental habitat integrity and health \citep{Higgs2010, Schliep2015}, geographical disease surveillance \citep{Jung2012} and seismic data \citep{Cameletti2017}. Spatio-temporal models for ordinal data have been developed for online ratings data \citep{Linero2018}, injury severity from traffic accidents \citep{Liu2019}, air quality data \citep{Ip2024} and drought levels \citep{Erhardt2024}. However, the multivariate ordinal likelihood in Bayesian models typically requires Markov chain Monte Carlo (MCMC) simulations \citep{Higgs2010, Schliep2015, Cameletti2017, Erhardt2024}. In the case of high dimensional latent variable settings that are common to spatial and spatio-temporal applications, MCMC approaches can be computationally demanding and require application-specific bespoke procedures to aid convergence.

The objective of this study is to propose the dynamic spatio-temporal sequential ordinal model (DSTSOM) for spatio-temporal applications. This spatio-temporal model for ordinal data can be fit using standard procedures developed for Bayesian spatio-temporal models of univariate data, for example, using the popular integrated nested Laplace approximation software package \citep[INLA;][]{Rue2009, Lindgren2015, Rue2017}. An approximate implementation of the proportional odds cumulative ordinal model available in INLA  uses an approximated Dirichlet prior \citep{Martinez-Minaya2024}. This approximation is required because the cumulative model for ordinal data relies on strictly ordered parameters that define the cutoff levels for the ordinal categories \citep{McCullagh1980}. The sequential ordinal model instead considers the transition probabilities from one level to another, which is a natural approach whenever the quantity of interest passes sequentially through categories \citep{Fahrmeir1994, Tutz2012}. The sequential ordinal model, unlike a cumulative ordinal model, does not impose strict ordering of cut levels and so simplifies estimation. The sequential ordinal model is shown particularly well-suited for spatial and spatio-temporal applications because its likelihood is equivalent to a univariate binary model conditional on covariates, coefficients and random effects. The widely applied separable space-time model developed for univariate data \citep[e.g.,][]{Cressie2011, Blangiardo2015} is shown to straightforwardly address multivariate ordinal data within the INLA framework.

The proposed dynamic spatio-temporal sequential ordinal model is demonstrated on an application from vegetation ecology.  Vegetation ecology often relies on expert assessments of plant cover using various modified versions of the Braun-Blanquet score \citep{Braun-Blanquet1932}. Braun-Blanquet scores constitute ordinal data and should  not be analysed as a numeric score \citep{Podani2006}. The vast majority of vegetation scores in a recently compiled global vegetation database use ordinal scoring with two--thirds based on modified Braun-Blanquet scores \citep{Bruelheide2019}, and the number of papers that rely on Braun-Blanquet scoring systems is rapidly growing \citep{Ivanova2024}. Here, ordinal observations of weeds in the Australian Capital Territory (ACT) based on modified Braun-Blanquet scores are considered. Weed distribution has important spatial and temporal dimensions that must be accounted for by specific weed management activities undertaken to control weed distribution and limit their negative environmental and economic  impacts. Spatial dependence is anticipated due to unobserved covariates, such as fine-scale habitat characteristics, that lead to greater similarity among nearby locations compared to those further apart. Both spatial and temporal dependence are essential considerations for accurately predicting weed distributions and evaluating the effectiveness of control measures as invasive weed populations grow and disperse through time. The invasive weeds ordinal data of the ACT are described as a modified and aggregated version of the Braun-Blanquet scores in Section \ref{sec:data}. The DSTSOM enables robust and accessible prediction of the abundance and spread of invasive alien plant species, thereby facilitating more effective targeting and evaluation of invasive weed control efforts.

The contributions of the paper are as follows. The  dynamic spatio-temporal sequential ordinal model (DSTSOM) is defined in Section \ref{sec:DSTSOM}. This ordinal model formulation extends dynamic generalised linear models to account for the multivariate ordinal response data in spatio-temporal models. The spatial autoregressive structure of the random effects is related to a commonly used convention in ordinal GLMs where the signs of covariates with global effects are reversed. The  implications of the sequential ordinal model and the global effects structure are interpreted for dynamic growth models of biological populations in Section \ref{sec:Gompertz}. A special case of the DSTSOM that is equivalent to a spatio-temporal cumulative ordinal proportional hazards model is shown in Section \ref{sec:prophaz}. This model has the useful property of collapsibility that maintains the interpretability of regression coefficients across aggregated ordinal categories. The DSTSOM allows estimation and prediction in INLA as with univariate spatio-temporal models (Section \ref{sec:INLA}). The approach is demonstrated on the invasive weeds application in Section \ref{sec:application}, where the incorporation of spatio-temporal dependence assists interpretation and prediction of the dynamic biological populations. Section \ref{sec:discussion} summarises contributions and outlines extensions.

\section{Ordinal Invasive Weeds Data}\label{sec:data}

The Australian Capital Territory (ACT) through its Biosecurity Strategy 2016--2026 \citep{ACT2016} seeks to prevent new weed infestations, minimise impact from weeds, and enhance weed management capacity. In this paper, weed monitoring and routine control activity (e.g., herbicide, manual removal, mowing) data from the ACT for the years 2019–-2023 were investigated for three perennial invasive grasses and a perennial herb species. The two related species, Chilean needle grass (\textit{Nassella neesiana}) and serrated tussock (\textit{Nassella trichotoma}), are classified as Weeds of National Significance (Invasive Plants and Animals Committee, 2016). A third grass species, African lovegrass (\textit{Eragrostis curvula}) and the perennial herb, Saint John’s wort (\textit{Hypericum perforatum}), are both considered high-threat invasive plants due to their impacts on rangeland productivity and the biodiversity of temperate grasslands and grassy woodlands across southern Australia \citep{CSIS2024, CSIS2024SJW}.

\begin{figure}
	\centering
	\includegraphics[width=1\textwidth]{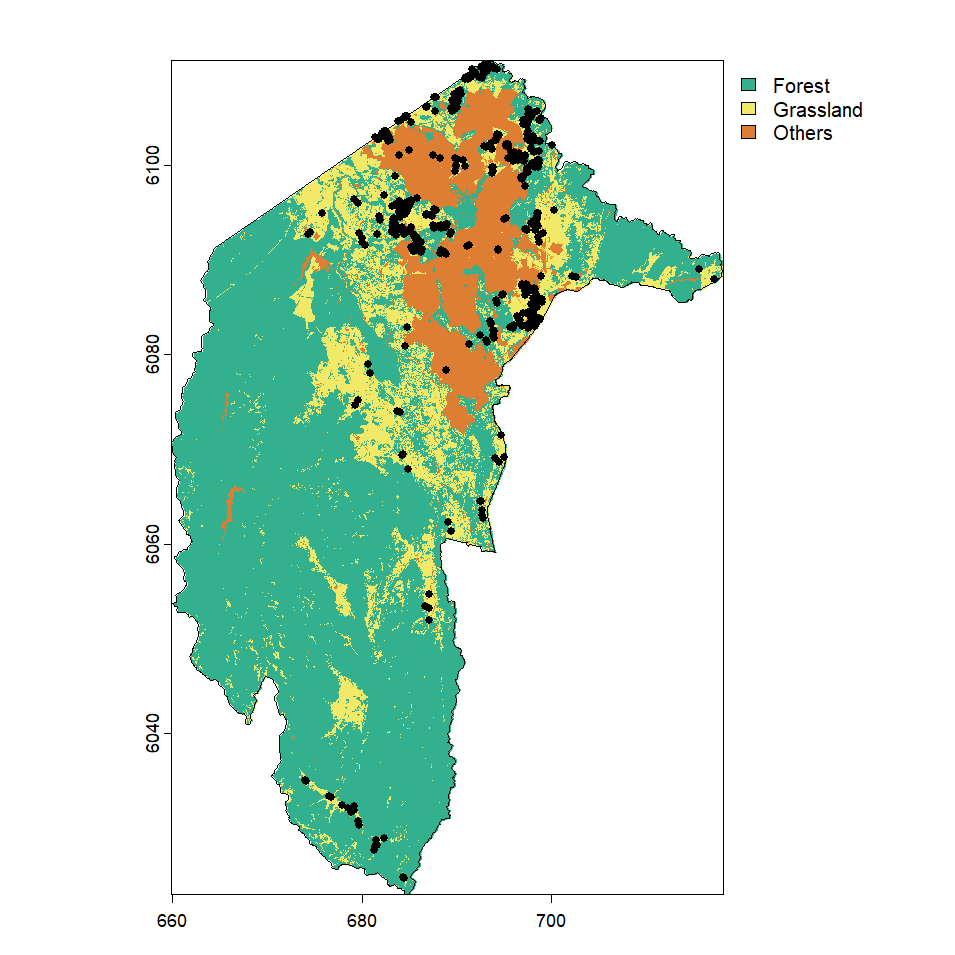}
	\caption{Map of study area showing forest habitat in green, grassland in yellow, and other habitats such as urban areas in orange within the Australian Capital Territory (ACT). Black points show site locations with observations. Observations and predictions in this study were restricted to forest and grassland habitats located inside the ACT. The axes show easting and northing coordinates expressed in units of kilometres. Based on data supplied by the ACT Government. \href{https://actmapi-actgov.opendata.arcgis.com/datasets/ACTGOV::actgov-border/about}{ACT border data} and \href{https://actmapi-actgov.opendata.arcgis.com/datasets/ACTGOV::actgov-vegetation-map-2018/about}{ACT vegetation data} used under \href{https://creativecommons.org/licenses/by/4.0/}{CC BY 4.0}.}\label{fig:studyarea}
\end{figure}

The geostatistical weeds monitoring data were assessed in the field and collated by the ACT using a modified Braun-Blanquet system \citep[see Table 1 in the Supplementary Information provided as a vignette to the supporting \textsf{R} package \textbf{DSTSOM},][]{DSTSOM2025}. Higher scores that correspond with high percent cover were rare for these species that are actively surveilled for control in the ACT \citep{SAGC2023}. The observations were therefore collapsed to an ordinal score that i) aggregated percent cover greater than 5\% and ii) added a lowest level category for species that were not recorded as present at an observation location (Table \ref{tab:data}). ACT control data were also available that provided the days and locations of targeted species-specific herbicide application. Between 181 and 334 observations were observed in each year between 2019 and 2023 (Supplementary Information). The 1,299 observations were inhomogeneously distributed through the ACT (Figure \ref{fig:studyarea}) because scoring of weeds by the modified Braun-Blanquet system occurs within fixed monitoring plots. These plots are located in designated  areas of the ACT that have high conservation value and are actively managed  to offset potential environmental impacts from development activities located elsewhere. Figure \ref{fig:studyarea} shows the locations of the monitoring plots within the combination of all woodland and forest vegetation communities, henceforth denoted as ``forest'' habitat, grassland habitat and other habitats such as urban areas (see Supplementary Information). The size of each monitoring plot is 400 square metres. All native and exotic vascular plant species present within each monitoring plot were identified to the species-level where possible, or genus-level for a few species where morphological species identification is difficult, and given a modified Braun-Blanquet score.  Monitoring effort was focused during spring to maximise the chances of observing a species, if present, because many species are more likely to show above ground or flower during spring. Monitoring effort in each year depended on prioritised time and resources available and was independent of weed control activities. The plots are surveyed to track floristic composition over space and time as an indicator of ecosystem condition to inform management decisions. Statistical modelling of the ordinal monitoring data for the invasive weeds must address spatio-temporal dependence.

\begin{table}
	\centering
	\caption{Ordinal categories were based on aggregated levels of the modified Braun-Blanquet scoring system used within the Australian Capital Territory for actively managed species of invasive weeds.}\label{tab:data}
	\begin{tabular}{|c|l|}
		\hline
		Ordinal Level & Description \\
		\hline
		1 & $0$\% cover and no individuals observed \\
		2 &	$<5$\% cover and solitary ($<4$ individuals) \\
		3 &	$<5$\% cover and few (4-15 individuals)\\
		4 &	$<5$\% cover and numerous/scattered ($>15$ individuals)\\
		5 &	$\geq5$\% cover\\
		\hline
	\end{tabular}
\end{table}

\section{Dynamic Spatio-Temporal Sequential Ordinal Models}\label{sec:DSTSOM}

\subsection{Dynamic Spatio-Temporal Generalised Linear Models}

Let  $y_i(\bm{s}_i, t)$ denote an observation at time $t$ and location with coordinates $\bm{s}_i$, for example, a spatial location with $\bm{s}_i \in \mathbb{R}^2$. The observations at time $t$ for $i = 1, \ldots, n(t)$ follow an exponential family in canonical form
\begin{equation}
	f(y_i(\bm{s}_i, t) \mid \eta_i(t), \psi) = \exp\left\{[y_i(\bm{s}_i, t)\eta_i(t) - b\left(\eta_i(t)\right)]/a(\psi) + c(y_i(\bm{s}_i, t), \psi)\right\}, \label{eq:GLM}
\end{equation}
for known functions $a(\cdot)$, $b(\cdot)$ and $c(\cdot)$ with conditional expectation $\mathbb{E}[y_i(\bm{s}_i, t) \mid \eta_i(t), \psi] = \mu_i(t)$ and monotonic link function $g$ that defines the linear predictor $\eta_i(t) = g(\mu_i(t)) \in \mathbb{R}$ of a generalised linear model \citep[GLM;][]{McCullagh1989}. Let $\bm{u}(t) = [u(\bm{s}_1, t) \ldots, u(\bm{s}_k, t)]^\top$ denote a $k$-dimensional vector of random effects at  time $t$ that follows a mean zero multivariate normal distribution with $k \times k$  covariance matrix $\bm{H}$ defined by the Mat{\'e}rn covariance function,
\begin{equation*}
	H_{ij} = \textrm{Cov}\left(u(\bm{s}_i, t), u(\bm{s}_j, t)\right) = \frac{\sigma^2}{\Gamma(\lambda)2^{\lambda - 1}}\left(\theta ||\bm{s}_i - \bm{s}_j ||\right)^\lambda K_\lambda(\theta ||\bm{s}_i - \bm{s}_j ||), \quad i, j \in \{1, \ldots, k\},
\end{equation*} 
where $||\cdot||$ is the Euclidean norm and $K_\lambda(\cdot)$ is the modified Bessel function of second kind and order $\lambda$. The scaling parameter $\theta$ sets the effective range $r = \sqrt{8 \lambda} / \theta$ as the Euclidean distance where the correlation is 0.13 for all $\lambda > 1/2$ \citep{Lindgren2015}. The dynamic spatio-temporal GLM is formulated by
\begin{align}
	y_i(\bm{s}_i, t) &\sim f\left(\mu_i(t), \psi\right), \quad i = 1, \ldots, n(t), \quad t = 1, \ldots, T, \label{eq:eta_tau} \\
	{g}\left({\mu}_i(t)\right) &= {\eta}_i(t), \quad i = 1, \ldots, n(t), \quad t = 1, \ldots, T,\nonumber\\ 
	\bm{\eta}(t) &= \bm{X}(t)\bm{\beta} + \bm{W}(t)\bm{u}(t), \nonumber\\
	\bm{u}(t) &= \bm{M}\bm{u}(t - 1) + \bm{\epsilon}(t), \quad \bm{\epsilon}(t) \sim N(\bm{0}, \bm{H}), \quad t = 2, \ldots, T,  \nonumber\\
	\bm{u}(1) &\sim N(\bm{0}, \bm{\Omega}),\nonumber
\end{align}
where the $p \times 1$ vector $\bm{\beta}$ are the  coefficients for the $n(t) \times p$ covariate matrix $\bm{X}(t)$. The $n(t) \times v$ matrix $\bm{W}(t)$ maps the random effects $\bm{u}(t)$ to the linear predictor at each time $t$ to allow for both missing observations and repeat observations at the same location. Eq. \eqref{eq:eta_tau} is an example of a dynamic generalised linear model \citep{West1997, Gelfand2003}, where the random effects follow an order-one autoregressive process with $v \times v$ state transition matrix $\bm{M}$. The $k$-dimensional vector of process innovations, $\bm{\epsilon}(t)$, accounts for process uncertainty in the latent autoregressive process at time $t$. 

Eq. \eqref{eq:eta_tau} is a flexible model for univariate responses that encapsulates, as special cases, a spatial GLM  if  $\bm{\Omega} = \bm{H}$ and $t = 1$, a longitudinal time series model if $\bm{H}$ is non-spatial, and a GLM without random effects if $\bm{W}(t) = \bm{0}$ for all $t$. In general, the spatio-temporal covariance function induced by the spatial autoregressive process model for $\bm{u}(t)$ is non-separable, for example, given diagonal $\bm{M}$ \citep{Cressie2011}. A separable spatio-temporal GLM \citep{Gelfand2003} is attained in Eq. \eqref{eq:eta_tau} if the space-time latent process uses 
\begin{equation}
	\bm{M} = \rho \bm{I}, \quad \rho \in (-1, 1), \quad \bm{\Omega} = \bm{H}/(1 - \rho^2),\label{eq:AR}
\end{equation}
where $\bm{I}$ is a $k \times k$ identity matrix and $\rho$ is a shared autocorrelation parameter \citep{Blangiardo2015}. The induced spatial and temporal covariance function for the random effects is then a separable product of a spatial and a temporal process that flexibly models spatial and temporal dependence \citep{Cameletti2013}.

\subsection{Cumulative and Sequential Ordinal Models}

Instead of the univariate GLM formulation of Eq. \eqref{eq:GLM} in the first line of Eq. \eqref{eq:eta_tau}, consider the case where observations are multivariate counts at location $\bm{s}_i$ and time $t$ from $C$ categories. Given the total count $m_i(t) \equiv m(\bm{s}_i, t)$ for the $i$\textsuperscript{th} observation at location $\bm{s}_i$ and time $t$ with $i = 1, \ldots, n(t)$, let the counts for the first $q = C - 1$ categories be denoted by the $q$-dimensional vector $\bm{y}_i(t) \equiv [y_{i1}(\bm{s}_i, t), \ldots, y_{iq}(\bm{s}_i, t)]^\top$  that follows the multinomial distribution,
\begin{align}
	P\left(\bm{y}_i(t) \mid \bm{\pi}_{i}(t), m_i(t)\right)=& \frac{m_i(t)!}{\left[m_i(t) - \sum_{c=1}^q y_{ic}(\bm{s}_i, t)\right]!\prod_{c=1}^q y_{ic}(\bm{s}_i, t)!}\label{eq:multiGLM}\\ 
	&\times \left(\prod_{c = 1}^q \pi_{ic}(t)^{y_{ic}(\bm{s}_i, t)}\right) \left(1 - \sum_{c = 1}^q \pi_{ic}(t) \right)^{m_i(t) - \sum_{c = 1}^q y_{ic}(\bm{s}_i, t)},\nonumber
\end{align}
where $\mathbb{E}[y_{ic}(\bm{s}_i, t) \mid \mu_{ic}(t), m_i(t)] =  m_i(t)\pi_{ic}(t)$ and the probabilities $\bm{\pi}_{i}(\bm{s}_i, t) \equiv \bm{\pi}_{i}(t) = [\pi_{i1}(t), \ldots, \linebreak \pi_{iq}(t)]^\top$ are such that $0 < \pi_{ic}(t) < 1$ for $c = 1, \ldots, q$ with $\sum_{c = 1}^q \pi_{ic}(t)  < 1$. If the $C$ categories are ordered such that $c_j < c_{j + 1}$ for $c = 1, \ldots, q$, then the multivariate observation $\bm{y}_i(t)$ in Eq. \eqref{eq:multiGLM} is ordinal.

Cumulative and sequential ordinal models are two common approaches used for ordinal multivariate GLMs \citep{Fahrmeir1994, Tutz2012}. Without loss of generality, let $z_i(t) \in \{1, \ldots, C\}$  denote the observed category for a single observation ($m_i(t) = 1$) at location $\bm{s}_i$ and time $t$, and for $c = 1, \ldots, q$ define the corresponding linear predictor $\eta_{ic}(t) = g(\pi_{ic}(t))$ with smoothly increasing link function $g: (0, 1) \rightarrow \mathbb{R}$. The cumulative ordinal model is defined by
\begin{equation}
	P\left(z_{i}(t) \leq c \mid \eta_{ic}(t)\right)\label{eq:cumul}
	= \gamma_{ic}(t) = g^{-1}\left(\eta_{ic}(t)\right), \quad \eta_{i(c - 1)}(t) < \eta_{ic}(t), \quad c = 1, \ldots, q, 
\end{equation}
with $\eta_{i0}(t) = -\infty$. The cumulative model probabilities by Eq. \eqref{eq:multiGLM} are $\pi_{ic}(t) = \gamma_{ic}(t) - \gamma_{i(c - 1)}(t)$ for $c = 1, \ldots, q$ and $i = 1, \ldots, n(t)$. The alternative sequential ordinal model is defined by
\begin{equation}
	P\left(z_{i}(t) = c \mid z_i(t) \geq c, \eta_{ic}(t)\right) = \delta_{ic}(t) =  g^{-1}\left(\eta_{ic}(t)\right), \quad c = 1, \ldots, q, \quad i = 1, \ldots, n(t). \label{eq:seq_mult}
\end{equation}
The sequential ordinal model does not impose the strict ordering of the linear predictors across the first $q$ categories mandated by the cumulative ordinal model in Eq. \eqref{eq:cumul}. By Eq. \eqref{eq:multiGLM}, the sequential ordinal model probabilities for $i = 1, \ldots, n(t)$ are
\begin{equation}
	P(z_i(t) = c \mid \bm{\pi}_i(t), m_i(t) = 1) = \pi_{ic}(t) = \delta_{ic}(t) \prod_{\tilde{c} = 1}^{c - 1} \left(1 - \delta_{i\tilde{c}}(t)\right), \quad c = 1, \ldots, q,\label{eq:ord_like}
\end{equation}
where $\prod_{r = 1}^{0} \emptyset = 1$ is the empty product for the case $c = 1$.

\subsection{Likelihood for Spatial Dynamic Sequential Ordinal Models}

Whereas Eq. \eqref{eq:eta_tau} is developed for a univariate response, the multinomial observation model suggested by Eq. \eqref{eq:multiGLM} for the ordinal models of Eqs. \eqref{eq:cumul} and \eqref{eq:seq_mult} is multivariate. The likelihood of a sequential ordinal GLM, however, can be recast as a binary regression model \citep{Armstrong1989, Berridge1991, Bender2000, Tutz2012}. Similar application to Eq. \eqref{eq:ord_like} accommodates the sequential ordinal model into the spatio-temporal GLM of Eq. \eqref{eq:eta_tau} as follows. 

Let $z_j(t) \equiv z_j(\bm{s}_j, t)  \in \{1, \ldots, C\}$ for $j = 1, \ldots, J(t)$ denote the $J(t)$ ordinal observations available at time $t$. Define the $n(t)$-dimensional binary observation vector 
\begin{equation*}
	\bm{y}(t) = [\bm{y}_1^\top(t), \ldots, \bm{y}_{J(t)}^\top(t)]^\top,
\end{equation*}
that is scored for categories $c = 1, \ldots, q = C - 1$, 
\begin{equation*}
	\bm{y}_j(t) = [y_{1 \mid j}(t), \ldots, y_{\zeta_j(t) \mid j}(t)]^\top, \quad \zeta_j(t) = \min(z_j(t), q),    \quad j = 1, \ldots, J(t),
\end{equation*}
with
\begin{equation*}
	y_{\tilde{c} \mid j}(t) = 
	\begin{cases}
			1 & \textrm{if } \tilde{c} = z_j(t) \\
			0 & \textrm{otherwise}
		\end{cases},
	\quad \tilde{c} = 1, \ldots, \zeta_j(t).
\end{equation*}
The above definition forms the vector $\bm{y}(t)$ of dimension $n(t) = \sum_{j = 1}^{J(t)} \zeta_j(t)$ in Eq. \eqref{eq:eta_tau}. Define the associated $n(t)$-dimensional linear predictor $\bm{\eta}(t) = [\bm{\eta}_1^\top(t), \ldots, \bm{\eta}_J^\top(t)]^\top$ for the sequential ordinal model, where $\bm{\eta}_j(t) = [\eta_{1 \mid j}(t), \ldots, \eta_{\zeta_j(t) \mid j}(t)]^\top$  with conformally partitioned matrices $\bm{X}(t)$ and $\bm{W}(t)$ in Eq. \eqref{eq:eta_tau} such that
\begin{equation}
	P\left(z_{j}(t) = c \mid z_j(t) \geq c, \eta_{c \mid j}(t)\right) = \delta_{c \mid j}(t) = g^{-1}\left(\eta_{c \mid j}(t)\right), \quad c = 1, \ldots, \zeta_j(t), \quad j = 1, \ldots, J(t).\label{eq:zeta}
\end{equation}
Here, with the linear predictor $\eta_{c \mid j}(t) = \bm{x}_{c \mid j}^\top(t)\bm{\beta} + \bm{w}_{c \mid j}^\top(t)\bm{u}(t)$ defined for $c = 1, \ldots, q$, it is evident that 
\begin{equation*}
	P\left(z_{j}(t) = C \mid z_j(t) \geq q, \eta_{q \mid j}(t)\right) = 1 - \delta_{q \mid j}(t).
\end{equation*}
Then, from Eq. \eqref{eq:ord_like}, the contribution of the $j$\textsuperscript{th} observation to the likelihood of the sequential ordinal model arises from a truncated multinomial distribution that for $j = 1, \ldots, J(t)$ is equivalent to the binary observation model,
\begin{align}
	P(z_j(t) \mid \bm{\pi}_{j}(t)) &= 
	\prod_{\tilde{c} = 1}^{\zeta_j(t)} \delta_{\tilde{c} \mid j}(t)^{y_{\tilde{c} \mid j}(t)} \left(1 - \delta_{\tilde{c} \mid j}(t)\right)^{1 - {y}_{\tilde{c} \mid j}(t)} \label{eq:bin01}\\
	&= \prod_{\tilde{c} = 1}^{\zeta_j(t)} g^{-1}\left(\eta_{\tilde{c} \mid j}(t)\right)^{y_{\tilde{c} \mid j}(t)} \left[1 - g^{-1}\left(\eta_{\tilde{c} \mid j}(t)\right)\right]^{1 - {y}_{\tilde{c} \mid j}(t)}.\nonumber
\end{align}
From Eq. \eqref{eq:eta_tau}, let $\bm{y} = [\bm{y}^\top(1), \ldots, \bm{y}^\top(T)]^\top$ and  $\bm{\eta} = [\bm{\eta}^\top(1)\top, \ldots, \bm{\eta}^\top(T)]^\top$. The full likelihood is
\begin{equation*}
	P(\bm{y} \mid \bm{\eta}) = \prod_{t = 1}^T \prod_{j = 1}^{J(t)} \prod_{\tilde{c} = 1}^{\zeta_j(t)} g^{-1}\left(\eta_{\tilde{c} \mid j}(t)\right)^{y_{\tilde{c} \mid j}(t)} \left[1 - g^{-1}\left(\eta_{\tilde{c} \mid j}(t)\right)\right]^{1 - {y}_{\tilde{c} \mid j}(t)}
\end{equation*}
for the dynamic spatio-temporal sequential ordinal model (DSTSOM).

\subsection{Global Effects Structure in DSTSOM}

Ordinal GLMs distinguish between covariates with category-specific effects, which vary across both observations and ordinal categories, and  covariates with global effects that vary only across observations. The values of a covariate with global effects are invariant across the ordinal categories for each observation. Allowing the covariates of a linear predictor to vary across categories can introduce too much flexibility and too many parameters in ordinal data analyses, and for this reason global effect structures are often used for ordinal GLMs \citep{McCullagh1989, Fahrmeir1994, Tutz2012}. In an ordinal GLM with global effects structure only the threshold parameters  vary among ordinal categories.

The global effects structure in the DSTSOM accounts for both covariates and random effects as follows. Let the first $q = C - 1$ entries of the $p$-dimensional coefficient vector $\bm{\beta}$ in the dynamic spatio-temporal sequential ordinal model relate to the  thresholds $\bm{\beta}_{1:(C - 1)} = [\beta_{1}, \ldots, \beta_{q}]^\top$ between the first $q$ categories, such that $\bm{\beta} = [\bm{\beta}_{1:q}^\top, \bm{\beta}_{C:p}^\top]^\top$ with coefficients $\bm{\beta}_{C:p} = [\beta_{C}, \ldots, \beta_{p}]^\top$ for the covariates. The linear predictor in Eq. \eqref{eq:zeta} for $j = 1, \ldots, J(t)$ then has the global effects structure
\begin{equation}
	\eta_{c \mid j}(t) = \beta_{c} - \bm{x}_{\ast \mid j}^\top(t)\bm{\beta}_{C:p} - \bm{w}_{\ast \mid j}^\top(t)\bm{u}(t), 
	\quad c = 1, \ldots, q,\label{eq:global_eta}
\end{equation}
where $\bm{x}_{\ast \mid j}(t) = [x_{j C}, \ldots, x_{j p}]^\top$ and $\bm{w}_{\ast \mid j}(t)$ denote the invariant global effects over the ordinal categories of the $j$\textsuperscript{th} observation. The sign reversal convention in Eq. \eqref{eq:global_eta} is used to stipulate that larger global effects increase the probabilities of higher level categories. 

The sign reversal convention is often used for covariates with global effects in ordinal GLMs \citep{McCullagh1980, Laara1985, Armstrong1989, Berridge1991, McCullagh1989, Tutz1991}, though not always \citep{Fahrmeir1994, Tutz2012}. It is also possible to use backwards sequential probabilities for the binary likelihood of ordinal GLMs \citep{Bender2000}. In the DSTSOM, the backwards approach would instead set $\delta_{c \mid j}(t) = P(z_j(t) = c \mid z_j \leq c, \eta_{c \mid j}(t))$ with the binary scoring of categories commensurately reversed in order. However, (forward) sequential probabilities are more commonly used in ordinal GLMs \citep{Armstrong1989, Berridge1991, Tutz1991, Fahrmeir1994, Tutz2012}. Here, the sign reversal convention for the (forward) sequential probabilities is applied to the covariates with global effects and extended also to the random effects in Eq. \eqref{eq:global_eta}. The sign reversal convention and its interpretation is illustrated for biological populations in the following section. 

\section{Gompertz Growth Model for Ordinal Data}\label{sec:Gompertz}

In population ecology, a Gompertz density dependent growth model for the logarithm of population sizes $\bm{\xi}(t)$ at time $t$ has the functional form
\begin{equation*}
	\log \bm{\xi}(t) = \bm{v} + \bm{B}\log\bm{\xi}(t - 1) + \bm{\epsilon}(t),
\end{equation*}
where $\bm{B}$ determines the strength of density dependence among populations, $\bm{v}$ sets the intrinsic population growth rates that would occur in the absence of density dependence ($\bm{B} = \bm{0}$), and $\bm{\epsilon}(t)$ captures process uncertainty and stochasticity. \citet{Ives2003}, for example, investigate properties of this dynamic structure for biological populations using vector autoregressive models with normal responses observed from a dynamic multispecies system.  \citet{Dennis2006} consider the case of single population time series with normal responses and \citet{Thorson2015} estimate a single species model with negative binomial response and spatio-temporal dependence in a spatial dynamic GLM. 

The intrinsic rates of growth in the Gompertz model may spatially vary in a way that can be modelled by covariates, for example, due to vital rates that vary with habitat type. Temporal-varying inputs may also be hypothesised to impact density independent rates of mortality, such as environmental change or management activities that alter population viability. Therefore an elaboration of the Gompertz growth model that allows spatio-temporal intrinsic rates of growth, $\bm{\nu}(t)$, to be described by covariates is implemented in the DSTSOM as follows. 

In Eq. \eqref{eq:eta_tau}, the covariates set the level of the interior predictor, but these can be shifted to instead reflect time-varying inputs into the latent trajectories using relations provided by \citet{Harvey1989} for dynamic linear models. Given the global effects structure of Eq. \eqref{eq:global_eta}, let the global effects contributed by the covariates to the $j$\textsuperscript{th} observation through Eq. \eqref{eq:eta_tau} be denoted by $\bm{w}_{\ast \mid j}^\top(t)\bm{\dot{\nu}}(t) = \bm{x}_{\ast \mid j}^\top(t)\bm{\beta}_{C:p}$. Let $\log \bm{\xi}(t) = \bm{u}(t) + \bm{\dot{\nu}}(t)$, where also $\bm{\dot{\nu}}(t) = \bm{\nu}(t) + \bm{M}\bm{\dot{\nu}}(t - 1)$ for $t = 1, \ldots, T$. The linear predictor of Eq. \eqref{eq:eta_tau} for category $c$ of observation $j$ with the sign reversal convention of Eq. \eqref{eq:global_eta} is then
\begin{equation}
	\eta_{c \mid j}(t) = \beta_c - \bm{w}_{\ast \mid j}^\top \log \bm{\xi}(t),\label{eq:bio_sign}
\end{equation}
with latent autoregressive structure 
\begin{equation}
	\log \bm{\xi}(t) = \bm{\nu}(t) +  \bm{M}\log \bm{\xi}(t - 1) + \bm{\epsilon}(t),\quad t = 1, \ldots, T,  \label{eq:Gompertz_AR1}
\end{equation}
given the initial conditions $\log \bm{\xi}(0) = \bm{u}(0)$ and $\bm{\dot{\nu}}(0) = \bm{0}$. The latent autoregressive structure of Eq. \eqref{eq:Gompertz_AR1} is equivalent to a Gompertz growth model for the population size with time-varying intrinsic rates of growth, $\bm{\nu}(t)$. Note that the sign of the state transition matrix $\bm{M}$ is unaffected by the sign reversal convention of Eq. \eqref{eq:bio_sign}. In a multispecies context \citep[sensu][]{Ives2003}, for example, the interpretation of competition, mutualism, commensalism, amensalism or predator-prey relationships that are defined by the signs of the off-diagonal entries of $\bm{M}$ is unaffected. 

In the same manner as the global effects structure of Eq. \eqref{eq:global_eta}, the sign reversal on the population size in Eq. \eqref{eq:bio_sign} is a convention that conforms with the interpretation where a larger population size increases the probability of higher level categories.  Note that increasing global effects modelled by the covariates lead to increased intrinsic rates of growth given the sign reversal convention. By Eqs. \eqref{eq:zeta} and \eqref{eq:bio_sign}, the conditional probability that the population is observed at level $c$ given that the level is at $c$ or above is 
\begin{equation*}
	P\left(z_{j}(t) = c \mid z_j(t) \geq c, \eta_{c \mid j}(t)\right) = \delta_{c \mid j}(t) = g^{-1}\left(\beta_c - \bm{w}_{\ast \mid j}^\top \log \bm{\xi}(t)\right).
\end{equation*}
The threshold parameters $\bm{\beta}_{1:q}$ thus determine the transition probabilities given the latent population size $\xi_{j}(t)$. The global structure of Eq. \eqref{eq:global_eta} within the sequential ordinal structure of the DSTSOM is therefore particularly well-suited for biological models of population growth, where the ordinal categories relate to the observed level of population size.  

The usefulness of the sign reversal convention is evident for the Gompertz growth model given the complementary log-log link function, 
\begin{equation}
	g\left(\delta_{c \mid j}\right) = \log\left[-\log\left( 1 - \delta_{c \mid j} \right)\right] = \eta_{c \mid j} \in \mathbb{R}.\label{eq:cloglog}
\end{equation}
Let also $\bm{w}_{\ast \mid j}^\top(t)$ indicate the latent population size for the $j$\textsuperscript{th} observation so that $\xi_j(t) = \bm{w}_{\ast \mid j}^\top(t) \log \bm{\xi}(t)$. With this choice, the sequential probabilities are modelled by
\begin{equation}
	\delta_{c \mid j}(t) = 1 - \exp\left[-d_{c \mid j}(t)\right],\label{eq:dd}
\end{equation}
where 
\begin{equation*}
	d_{c \mid j}(t) = \exp\left(\eta_{c \mid j}(t)\right) = \exp\left(\beta_c\right)\xi_j(t)^{-1}.
\end{equation*}
Eq. \eqref{eq:dd} has the same form as the probability of a non-zero response for a Poisson distribution with mean or intensity $d_{c \mid j}(t)$. The probability that the population size is observed at level $c$ increases with the threshold parameter $\beta_c$, whereas the probability of observations at levels greater than $c$, $P\left(z_{j}(t) > c \mid z_j(t) \geq c, \eta_{c \mid j}(t)\right) = 1 - \delta_{c \mid j}(t)$, increases with the population size $\xi_j(t)$. 

In a separable model with the specification of Eq. \eqref{eq:AR}, the latent Gompertz process model for the logarithm of the $k$\textsuperscript{th} population size has the readily interpretable form,
\begin{equation*}
	\log \xi_k(t) = v_k(t) + \rho \xi_k(t - 1) + \epsilon_k(t).
\end{equation*}
The parameter  $b = \rho - 1$ is the strength of density dependent growth of the population in the Gompertz model of population growth \citep{Dennis2006}. As the density dependence parameter $a$ approaches zero then the Gompertz model approaches a density independent model of exponential growth.

\section{Proportional Hazards Equivalence and Collapsibility}\label{sec:prophaz}

For certain model structures there is an equivalence between the cumulative and sequential model. In an ordinal GLM with the complementary log-log link function of Eq. \eqref{eq:cloglog} and covariates with global effects, the sequential model is equivalent to the popular grouped Cox model \citep{Laara1985} that is also known as the proportional hazards model \citep{McCullagh1980, McCullagh1989}. These latter models are examples of cumulative models. All cumulative models have the property of ``collapsibility'' \citep{Tutz2012}, which is a preferred property of ordinal models that holds if the interpretation of a parameter is unchanged after grouping ordinal categories together \citep{Tutz1991}. 

The equivalence and collapsibility properties hold for sequential ordinal models with complementary log-log link function and random effects that like the covariates are shared across all categories for each observation as in the global structure for the linear predictor in Eq. \eqref{eq:global_eta}. To see this, let the cumulative proportional hazards models defined by Eq. \eqref{eq:cumul} with  complementary log-log link  have for $\tilde{\eta}_{jc}(t) \equiv \tilde{\eta}_{c \mid j}(t)$ the global structure analogous to Eq. \eqref{eq:global_eta},
\begin{equation*}
	\tilde{\eta}_{c \mid j}(t) = \tilde{\beta}_{c} - \bm{x}_{\ast \mid j}^\top(t)\bm{\tilde{\beta}}_{C:p} - \bm{w}_{\ast \mid j}^\top(t)\bm{\tilde{u}}(t),\quad c = 1, \ldots, q.
\end{equation*}
The coefficient $\tilde{\beta}_0 = -\infty$ is fixed in the cumulative model. To show collapsibility for an ordinal response $\bm{\check{z}}(t)$ with aggregated categories derived from $\bm{z}(t)$, let also the subsets $S_l = \{k_{l - 1}, \ldots, k_l\}$ for $l = 1, \ldots, K$ with $k_0 = 0$ and $k_K = C$ for $1 < K \leq C$ form a partition of the response categories such that $S_a \cap S_b = \emptyset$ for $a \neq b$ and $\cup_{l = 1}^K S_l = \{1, \ldots, C\}$. The resulting model for $l = 1, \ldots, K - 1$ is
\begin{equation}
	P\left(\check{z}_{j}(t) \leq l \mid \bm{\check{\beta}}, \bm{\tilde{\beta}}_{K:(p - C + K)}, \bm{\tilde{u}}(t)\right) =  1 - \exp\left\{-\exp\left[\check{\beta}_{l} - \bm{x}_{\ast \mid j}^\top(t)\bm{\tilde{\beta}}_{K:(p - C + K)} - \bm{w}_{\ast \mid j}^\top(t)\bm{\tilde{u}}(t)\right]\right\},\label{eq:PH}
\end{equation}
where $\bm{\check{\beta}}$ is the vector of threshold parameters with dimension $K - 1$, and $\bm{\tilde{\beta}}$ is the vector of global effect coefficients with dimension $p - C + 1$. The case $1 < K < C$ illustrates the collapsibility of the proportional hazards model with respect to those coefficients related to covariates with global effects, and also the random effects that are global across all categories for the $j$\textsuperscript{th} observation. 

For equivalence, given Eq. \eqref{eq:PH} note that the optionally collapsed sequential ordinal model of Eq. \eqref{eq:seq_mult} with complementary log-log link function for globally structured $\eta_{jl}(t) = \eta_{l \mid j}(t)$ and $l = 1, \ldots, K - 1$ is
\begin{align*}
	\eta_{l \mid j}(t) &= \beta_l - \bm{x}_{\ast \mid j}^\top(t) \bm{\beta}_{K:(p - C + K)} - \bm{w}_{\ast \mid j}^\top(t)\bm{u}(t)\label{eq:equiv}\\
	&= \log \left[-\log \left(1 - P\left(\check{z}_j(t) = l \mid \check{z}_j(t) \geq l, \bm{{\beta}}, \bm{u}(t)\right)\right)\right]\nonumber.
\end{align*}
Expression in the form of the cumulative model would mean that the last line above is equal to
\begin{align*}	
	\log &\left[-\log \frac{1 - P\left(\check{z}_j(t) \leq l  \mid \bm{\check{\beta}}, \bm{\tilde{\beta}}_{K:(p - C + K)}, \bm{\tilde{u}}(t)\right)}{1 - P\left(\check{z}_j(t) \leq l - 1 \mid \bm{\check{\beta}}, \bm{\tilde{\beta}}_{K:(p - C + K)}, \bm{\tilde{u}}(t)\right)}\right] \nonumber \\
	&= \log\left[\exp\left(\check{\beta}_l\right) - \exp\left(\check{\beta}_{l - 1}\right)\right] - \bm{x}_{\ast \mid j}^\top(t) \bm{\tilde{\beta}}_{K:(p - C + K)} - \bm{w}_{\ast \mid j}^\top(t)\bm{\tilde{u}}(t).\nonumber
\end{align*}
The parameterisation
\begin{equation*}
	\beta_{l} = \log\left[\exp(\check{\beta}_{l}) - \exp(\check{\beta}_{l - 1})\right], \quad l = 1, \ldots, K - 1,
\end{equation*}
shows that the DSTSOM with global structure and complementary log-log link is equivalent to the proportional hazards model both with and without collapsed categories given the coefficients $\bm{\beta}_{K:(p - C + K)} = \bm{\tilde{\beta}}_{K:(p - C + K)}$ and random effects $\bm{u}(t) = \bm{\tilde{u}}(t)$.

\section{INLA Estimation}\label{sec:INLA}

Accommodation of spatio-temporal models with non-Gaussian models requires estimation procedures, such as INLA, that account for the increased dimensionality of random effects and computational complexity induced by non-normal likelihoods \citep{Cressie2011, Rue2017}. Software functionality for ordinal models in INLA was absent \citep{Schliep2015, Cameletti2017} until the recent addition of an approximation to the multivariate Dirichlet distribution \citep{Martinez-Minaya2024}. The Dirichlet prior approximation in INLA supports an implementation of the cumulative proportional hazards model. It is also possible to approximate the multinomial likelihood, as mentioned in the INLA documentation, by appealing to the kernel of a vector of Poisson random variables \citep[sensu][]{Baker1994}. Here, instead of the cumulative ordinal model of Eq. \eqref{eq:cumul}, the sequential ordinal model of Eq. \eqref{eq:seq_mult} is used to avoid the need for approximate models in the prior structure or likelihood of Eq. \eqref{eq:eta_tau}. As shown in Section \ref{sec:DSTSOM}, the recasting of the sequential ordinal model by means of the binary likelihood representation allows its incorporation into an effectively univariate estimation framework without need for approximation in an ordinal GLM. Univariate implementations of the dynamic spatio-temporal sequential ordinal model developed for INLA are then directly applicable to the multivariate ordinal data with this approach. \citet{Cameletti2013} and \citet{Blangiardo2015}, for example, investigate dynamical models in a space-time INLA application of Eq. \eqref{eq:eta_tau}, where $f$ is a normal distribution with variance $\psi$ and $\bm{g}$ is the identity function given the separable specification of Eq. \eqref{eq:AR}. \citet{Blangiardo2015} demonstrates estimation of this Bayesian separable first order autoregressive models in discrete time with INLA, along with non-Gaussian and multivariate extensions. 

The INLA approach for the univariate spatio-temporal dynamic model is also applicable to the DSTSOM model with the parameter specifications of Eq. \eqref{eq:AR} using instead a binary likelihood model and appropriate link function in  Eq. \eqref{eq:eta_tau}. The matrix $\bm{W}(t)$ in Eq. \eqref{eq:eta_tau} accounts for shared random effects among the binary likelihood contributions that are derived from the corresponding single ordinal observation. Therefore, although the ordinal model is multivariate,  the approach of Section \ref{sec:DSTSOM} for DSTSOMs does not use  INLA functionality that copies nearly identical random effects to different likelihoods \citep{Blangiardo2015, Krainski2019}, which requires the addition of a small noise term for computational reasons \citep{Martins2013, Gomez-Rubio2020}. Instead, the random effects shared among the binary observations that relate to a single ordinal observation are assigned by the projection matrix $\bm{A}$ that maps a mesh approximation of the random field to the observations available or predicted at specific sites and times. See, for example, \citet{Blangiardo2015, Krainski2019} for further implementation details of the projection matrix in INLA and its role in computing predictions by drawing joint samples from the posterior.  Note that the choice of spatial mesh used by an INLA implementation is linked to both the observations' locations and the spatial prior specification \citep{Righetto2020, Verdoy2021, Dambly2023}, and this link also holds for  dynamic spatio-temporal GLMs through the choice of spatial covariance function and the matrix $\bm{H}$ in Eq. \eqref{eq:eta_tau}. 

The sign reversal convention for models with global effects, as in Eq. \eqref{eq:global_eta}, is easily implemented in INLA by simply multiplying the global effect covariates with negative one. The correspondingly altered random effects $\bm{u}'(t)$ and process innovations $\bm{\epsilon}'(t)$ estimated in INLA are then accounted for in the Gompertz growth model (Section \ref{sec:Gompertz}) by reversing the sign of the initial population size $\log \bm{\xi}(0) = -\bm{u}'(0)$ and the innovations $\bm{\epsilon}(t) = \bm{\epsilon}'(t)$ for $t = 1, \ldots, T$ in Eq. \eqref{eq:Gompertz_AR1}. However, this step is obviously irrelevant if focus is only on posterior and predictive posterior estimates of the sequential probabilities rather than the latent population sizes that are only defined through the ordinal categorisation. Note that the sign of the state transition matrix $\bm{M}$ and the interpretation of density dependence in the Gompertz growth model is unaffected by the sign reversal convention for global effects in Eq. \eqref{eq:Gompertz_AR1}. A synthetic data example of a DSTSOM fit using INLA with the sign reversal convention is provided in the Supplementary Information.

\section{Application to Invasive Weeds of the ACT}\label{sec:application}

This section begins with exploratory analysis and description of the ordinal invasive weeds data (Section \ref{sec:data}), then introduces the model structure and prior parameterisation before presenting posterior estimates and predictive posterior spatio-temporal predictions of the ordinal categories of each species. The active management of the weed species likely reduces interspecific competition for resources and the magnitude of statistical dependencies among species. An exploratory analysis of the ordinal scores, for example, found only weak Kendall rank correlations that ranged between 0.00 and 0.17 across all pairwise interspecific comparisons. However, it is a different story within a species, where strong temporal dependence is expected. The abundance of a perennial weed at a site in one year is likely to persist in subsequent years without management intervention. The species also spread using different strategies such as wind-assisted, human-assisted and animal-assisted dispersal as well as vegetative growth \citep{CSIS2024, CSISCNG2024, CSIS2024SJW, CSIS2024Serr} that can induce spatial dependence \citep{Bahn2008}. Spatial dependence within species may arise either from fine-scale habitat characteristics (e.g. soil type, disturbance history) or by local seed dispersal and vegetative growth. Human-mediated dispersal could be modelled by proximity of sites to trails and roads with vehicle traffic \citep{Christen2006, Ansong2013}, although limited access may also result in higher weed cover due to reduced control effort. These ecological dynamics vary across species, reflecting differences in life history traits and habitat preferences \citep{Zeil-Rolfe2024}, such as varying affinities for forest versus grassland ecotypes (Figure \ref{fig:studyarea}). As such, the statistical modelling approach must account for species-specific control history, spatial predictors, and both spatial and temporal dependence.

\subsection{Model Structure}

The growth of weed abundance and percent cover occurs through time leading to higher level ordinal categories in the Braun-Blanquet scoring of Table \ref{tab:data}. Therefore the sequential ordinal model representation that models transitions between ordinal levels in Eq. \eqref{eq:seq_mult} was chosen for each species as described in Section \ref{sec:data}. The underlying growth process for the ACT invasive weeds (Section \ref{sec:data}) occurs across years and so the discrete time DSTSOM (Section \ref{sec:DSTSOM}) is employed.  The derived ordinal levels are interpreted as an ordinal proxy to a level of weed population size (Section \ref{sec:Gompertz}), and so the added flexibility that would be provided by modelling category specific covariates and random effects was undesirable compared to a global structure for both covariates and random effects.  As described in Section \ref{sec:data}, the sparse data observations used an ordinal scoring system with aggregated levels and so the property of collapsibility was therefore preferred. Therefore the implemented model structure used the complementary log-log link function (Section \ref{sec:prophaz}). The sparse observations of low percent cover (Section \ref{sec:data})  suggested that a separable space-time autoregressive structure independently specified for each species in Eq. \eqref{eq:AR} was a reasonably flexible model for the spatio-temporal covariance function of the random effects.

Spatial covariates included a forest versus grassland factor, $F(\bm{s}_i)$, and a continuous log transformed access covariate, $L(\bm{s}_i)$, defined by the distance to roads and trails in kilometres (Section \ref{sec:data}) for a location $\bm{s}_i$. A linear trend for year was included to account for the general environmental effect on weed abundance across 2019--2023. For weed control activities, let $d_i(\bm{s}_i , h)$ denote the duration since the last targeted control event for the weed species $h$ at time $t$ and location $\bm{s}_i$. The duration was considered in units of years but allowed to vary by day to accommodate intra-annual variation in control activities. A binary factor for the exposure of species $h$ to a weed control event was coded by $E_i(\bm{s}_i , h) = 1$ if $d_i(\bm{s}_i , h) > 0$ and $E_i(\bm{s}_i , h) = 0$. Together these spatio-temporal covariates allowed for a shift in weed population size following a control event and subsequent change in temporal trend. 

The contribution of the covariates to the linear predictor for category $c$ of the $i$\textsuperscript{th} observation at location $\bm{s}_i$ in year $t$ in the DSTSOM for species $h$ is then summarised by
\begin{equation*}
	\bm{x}_{c \mid i}^\top(t) = [cut(1), \ldots, cut(c), \ldots, cut(q),  E_i(\bm{s}_i , h), d_i(\bm{s}_i , h), t_i, F(\bm{s}_i), L(\bm{s}_i)],
\end{equation*}
where $cut(c)$ is a binary factor that for $c = 1, \ldots, q$ is equal to one for category level $c$ and zero otherwise. The associated coefficients are
\begin{equation*}
	\bm{\beta} = [\beta_{cut\, 1}, \ldots, \beta_{cut\, c}, \ldots, \beta_{cut\, q}, -\beta_{ctrl\, h}, -\beta_{d\, h}, -\beta_{year}, -\beta_{forest}, -\beta_{log\, access}]^\top.
\end{equation*}
Note that the coefficients of the global effects are reversed in sign, as in Eq. \eqref{eq:global_eta}, so that increases in the global effect covariates are associated with increased probability of an observation occurring at higher ordinal levels (Section \ref{sec:DSTSOM}). The parameter $\beta_{cut\ c}$ is a threshold parameter for category level $c = 1, \ldots, q$ in the sequential ordinal model. The impact of the control event for species $h$ has impact of magnitude $\beta_{ctrl\, h}$. If $\beta_{ctrl\, h} < 0$ ($\beta_{ctrl\, h} > 0$) then the weed control event  has a negative (positive) association with the density independent growth rate of the target weed species in the Gompertz population growth model (Section \ref{sec:Gompertz}). Note, however, that a causal negative impact of a control event may be masked by observation bias. For example, if sites with more abundant weeds are more likely to be actively targeted for control, then the statistical association between control events and weed population size could be negated or even switch sign relative to the causal impact of control. If $\beta_{d\, h}$ has contrasting sign to $\beta_{ctrl\, h}$ then the population initially returns towards its baseline trend after the control event. The coefficient $\beta_{year}$ shows a population decline if negative and a population increase trending over time if positive. Positive coefficients for $\beta_{forest}$ and $\beta_{log\, access}$ indicate greater weed abundance with forest habitat compared to grassland and increasing distances from roads and trails, respectively. 

In addition to the spatio-temporal model, an ordinal GLM without random effects ($M_1$) and a model with only spatial dependence ($M_2$) were also fit for each of the four species (Supplementary Information); the full separable spatio-temporal model is $M_3$. These models were evaluated to assess the influence of spatial and spatio-temporal dependence  on posterior estimates of the coefficients associated with the spatial and spatio-temporally varying covariates, including the effect of control events as captured by the parameters $\beta_{ctrl\ h}$ and $\beta_{d\ h}$. Predictive posterior medians and central 90\% credible intervals were estimated for each ordinal level and year from the full  model with spatio-temporal dependence for each species. The deviance information criterion \citep{Spiegelhalter2002} was also calculated for each model, where lower values suggest better fits.

\subsection{Prior Parameterisation of DSTSOM}

The prior for the spatial-temporal random effect standard deviation $\sigma$  of the Mat{\'e}rn covariance function  was adjusted for the sequential ordinal model using the following rationale. For a binary likelihood with a complementary log-log link, a marginal standard deviation of one would suggest a 95\% prior central credible interval that could span mean probabilities from about 0.05 to 0.95 for the binary response: This a priori variation ascribed to the random effect would, for example, arise if the linear predictor was otherwise held constant near $-0.90$. Equivalently, if $\sigma = 1$ then there is a 5\% chance that a random effect with mean zero exceeds $\pm 1.96$. However, the sequential ordinal model additionally imposes $q$ cut levels or threshold parameters between ordinal levels. If the marginal standard deviation is instead $\sigma = 1/q$, then there is a 5\% chance that a random effect with mean zero exceeds $\pm 1.96/q$. The marginal variation in the random effect was therefore adjusted to account for the number of ordinal categories. A penalised complexity prior \citep{Fuglstad2019} was specified such that $P(\sigma > 1/q) = 0.05$. 

For the effective range of the Mat{\'e}rn covariance function with order $\lambda = 1$, a penalised complexity prior set $P(r < 10~km) = 0.05$. The 10 km bound on the effective range $r$ was about 10\% of the maximum distance between observed locations (Figure \ref{fig:studyarea}).  For the spatial mesh specification in INLA, the cutoff that limits the size of the smallest triangles in the mesh were set to 500 metres, which is 1/20 of the 10 km probabilistic bound used in the penalised complexity prior for the spatial range. The inner maximum edge argument was increased two-fold relative to the cutoff to 1 km and the outer maximum edge argument was increased another ten-fold to 10 km.

For the temporal correlation parameter $\rho$, a penalised complexity prior specified $P(\rho > 0.50) = 2/3$, meaning that a priori it was envisioned that the autocorrelation parameter was twice as likely to be above $0.50$ as below. This choice reflected two considerations. First, the weeds species are actively managed and at low abundance such that density dependence in the Gompertz growth model should be relatively limited, and so the prior favours low values of density dependence (Section \ref{sec:Gompertz}). Second, the weed density in the current year depends on weed density in the year previous due to localised population growth, which suggests positive temporal autocorrelation. For stationarity, it was assumed that $-1 < \rho < 1$. The coefficients $\bm{\beta}$ were assigned broad mean zero independent normal priors with a large variance of 1000. 

\subsection{Results}

\begin{table}[!htbp]
	\centering
	\caption{Posterior estimates of parameters for the dynamic spatio-temporal sequential ordinal models applied to four species of actively managed invasive weeds.}\label{tab:spp}
	\small 
	\begin{tabular}{llrrr}
		\hline
		Species & Parameter & q0.025 & q0.50 & q0.975 \\ 
		\hline
\multirow{12}{*}{African lovegrass} & cut\_1 & 0.576 & 1.167 & 1.816 \\ 
& cut\_2 & $-$0.039 & 0.617 & 1.339 \\ 
& cut\_3 & 0.827 & 1.511 & 2.271 \\ 
& cut\_4 & 1.447 & 2.208 & 3.053 \\ 
& ctrl\_AGC & $-$0.389 & $-$0.030 & 0.325 \\ 
& d\_AGC & $-$0.025 & 0.063 & 0.151 \\ 
& year & 0.126 & 0.211 & 0.301 \\ 
& forest & $-$0.566 & $-$0.282 & $-$0.002 \\ 
& log\_access & $-$0.132 & $-$0.006 & 0.122 \\ 
& $r$ & 2.664 & 3.774 & 5.389 \\ 
& $\sigma$ & 1.094 & 1.408 & 1.806 \\ 
& $\rho$ & 0.981 & 0.998 & 1.000 \\ 
		\hline
\multirow{12}{*}{Chilean needle grass} & cut\_1 & 1.226 & 1.863 & 2.670 \\ 
& cut\_2 & $-$0.052 & 0.712 & 1.635 \\ 
& cut\_3 & 0.284 & 1.084 & 2.044 \\ 
& cut\_4 & 2.232 & 3.046 & 4.044 \\ 
& ctrl\_CNG & $-$0.327 & 0.381 & 1.080 \\ 
& d\_CNG & $-$0.152 & $-$0.024 & 0.104 \\ 
& year & $-$0.034 & 0.063 & 0.164 \\ 
& forest & $-$0.468 & $-$0.175 & 0.109 \\ 
& log\_access & $-$0.237 & $-$0.098 & 0.028 \\ 
& $r$ & 2.982 & 4.560 & 7.110 \\ 
& $\sigma$ & 0.879 & 1.188 & 1.597 \\ 
& $\rho$ & 0.944 & 0.990 & 0.999 \\ 
		\hline
\multirow{12}{*}{Saint John's wort} & cut\_1 & $-$1.097 & $-$0.232 & 0.642 \\ 
& cut\_2 & $-$1.387 & $-$0.509 & 0.383 \\ 
& cut\_3 & $-$0.506 & 0.371 & 1.263 \\ 
& cut\_4 & 2.298 & 3.191 & 4.116 \\ 
& ctrl\_StJW & 0.016 & 0.348 & 0.681 \\ 
& d\_StJW & $-$0.049 & 0.038 & 0.125 \\ 
& year & $-$0.057 & 0.165 & 0.366 \\ 
& forest & $-$0.456 & $-$0.246 & $-$0.038 \\ 
& log\_access & $-$0.199 & $-$0.096 & 0.009 \\ 
& $r$ & 4.658 & 6.410 & 8.864 \\ 
& $\sigma$ & 1.444 & 1.781 & 2.202 \\ 
& $\rho$ & 0.861 & 0.921 & 0.956 \\ 
		\hline  
\multirow{12}{*}{Serrated tussock} & cut\_1 & 0.884 & 1.364 & 1.895 \\ 
& cut\_2 & 1.040 & 1.580 & 2.180 \\ 
& cut\_3 & 1.402 & 2.006 & 2.672 \\ 
& cut\_4 & 2.501 & 3.218 & 4.008 \\ 
& ctrl\_St & 0.415 & 0.741 & 1.072 \\ 
& d\_St & $-$0.179 & $-$0.104 & $-$0.029 \\ 
& year & 0.114 & 0.211 & 0.315 \\ 
& forest & $-$0.336 & $-$0.114 & 0.109 \\ 
& log\_access & $-$0.107 & $-$0.004 & 0.099 \\ 
& $r$ & 2.533 & 3.903 & 6.188 \\ 
& $\sigma$ & 0.741 & 0.973 & 1.271 \\ 
& $\rho$ & 0.884 & 0.955 & 0.985 \\ 
		\hline
	\end{tabular}
\end{table}

\begin{table}[!htbp]
	\centering
	\caption{Deviance information criteria for four species of actively managed invasive weeds evaluated for alternative models of dependence: $M_1$, sequential ordinal GLM without random effects; $M_2$, spatial sequential ordinal model; and $M_3$, dynamic spatio-temporal sequential ordinal model.}\label{tab:dic}
	\normalsize 
	\begin{tabular}{lcr}
		\hline
		Species & Model & DIC \\ 
		\hline
\multirow{3}{*}{African lovegrass} & M1 & 1631 \\ 
& M2 & 1259 \\ 
& M3 & 1262 \\ 
		\hline
\multirow{3}{*}{Chilean needle grass} & M1 & 1348 \\ 
& M2 & 956 \\ 
& M3 & 959 \\ 
		\hline
\multirow{3}{*}{Saint John's wort} & M1 & 3444 \\ 
& M2 & 2740 \\ 
& M3 & 2625 \\ 
		\hline
\multirow{3}{*}{Serrated tussock} & M1 & 1785 \\ 
& M2 & 1520 \\ 
& M3 & 1507 \\ 
		\hline
	\end{tabular}
\end{table}

\begin{figure}
	\includegraphics[width=0.95\textwidth]{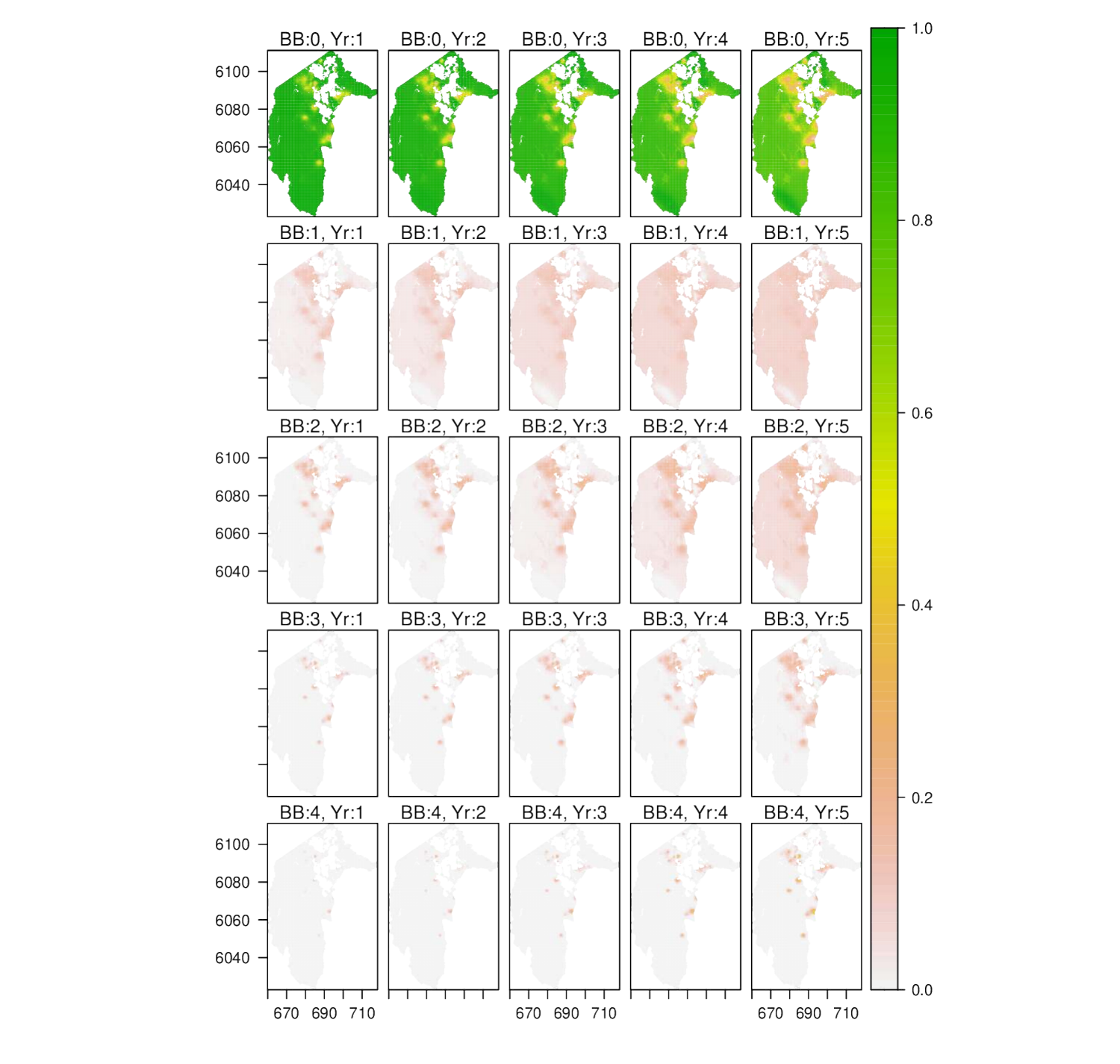}
	\caption{The predictive posterior median probability of the aggregated modified Braun-Blanquet scores for African lovegrass by years (columns) and scores (rows). Based on data supplied by the ACT Government. \href{https://actmapi-actgov.opendata.arcgis.com/datasets/ACTGOV::actgov-border/about}{ACT border data} and \href{https://actmapi-actgov.opendata.arcgis.com/datasets/ACTGOV::actgov-vegetation-map-2018/about}{ACT vegetation data} used under \href{https://creativecommons.org/licenses/by/4.0/}{CC BY 4.0}.}\label{fig:AGC}
\end{figure}

\begin{figure}
	\includegraphics[width=0.95\textwidth]{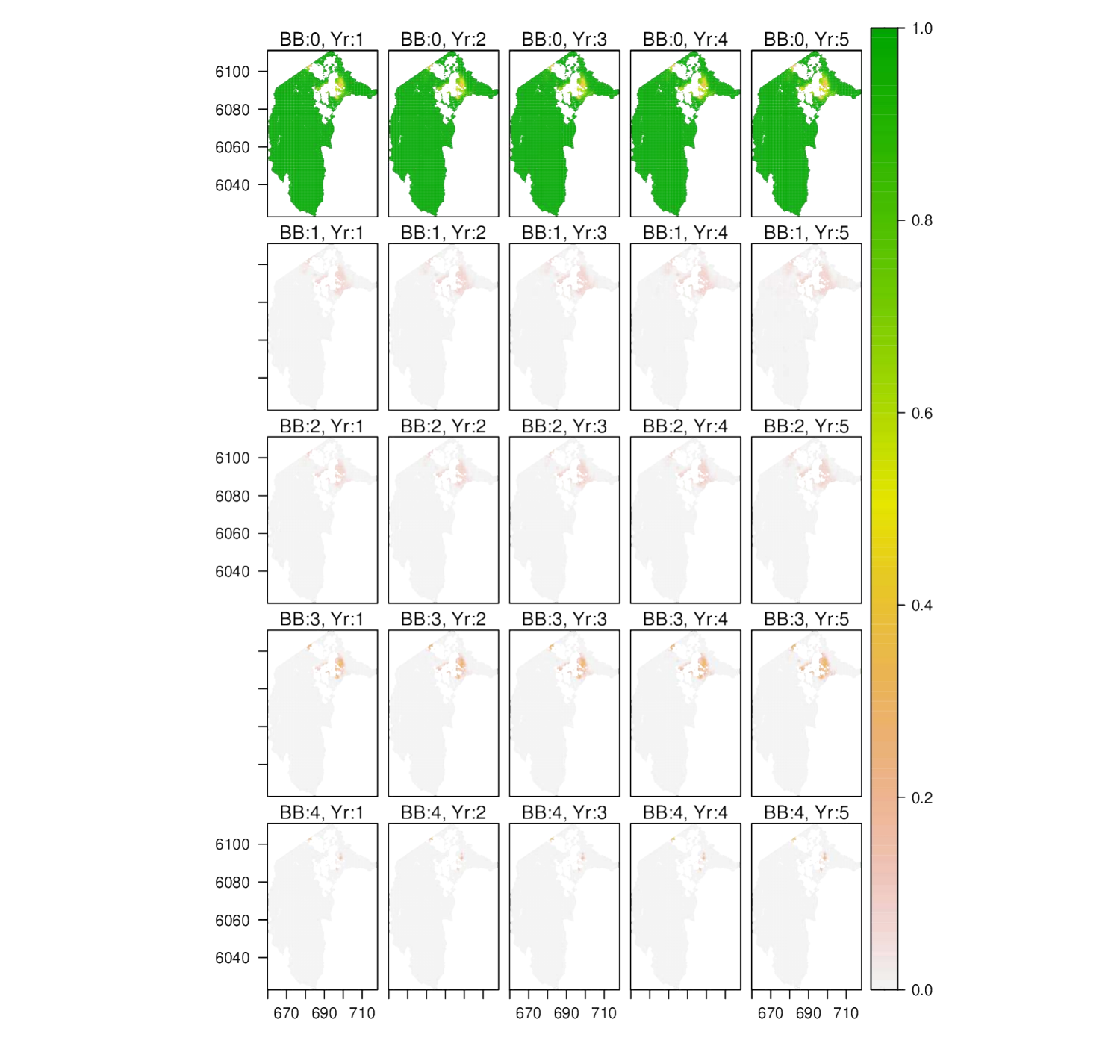}
	\caption{The predictive posterior median probability of the aggregated modified Braun-Blanquet scores for Chilean needle grass by years (columns) and scores (rows). Based on data supplied by the ACT Government. \href{https://actmapi-actgov.opendata.arcgis.com/datasets/ACTGOV::actgov-border/about}{ACT border data} and \href{https://actmapi-actgov.opendata.arcgis.com/datasets/ACTGOV::actgov-vegetation-map-2018/about}{ACT vegetation data} used under \href{https://creativecommons.org/licenses/by/4.0/}{CC BY 4.0}.}\label{fig:CNG}
\end{figure}
	
\begin{figure}
	\includegraphics[width=0.95\textwidth]{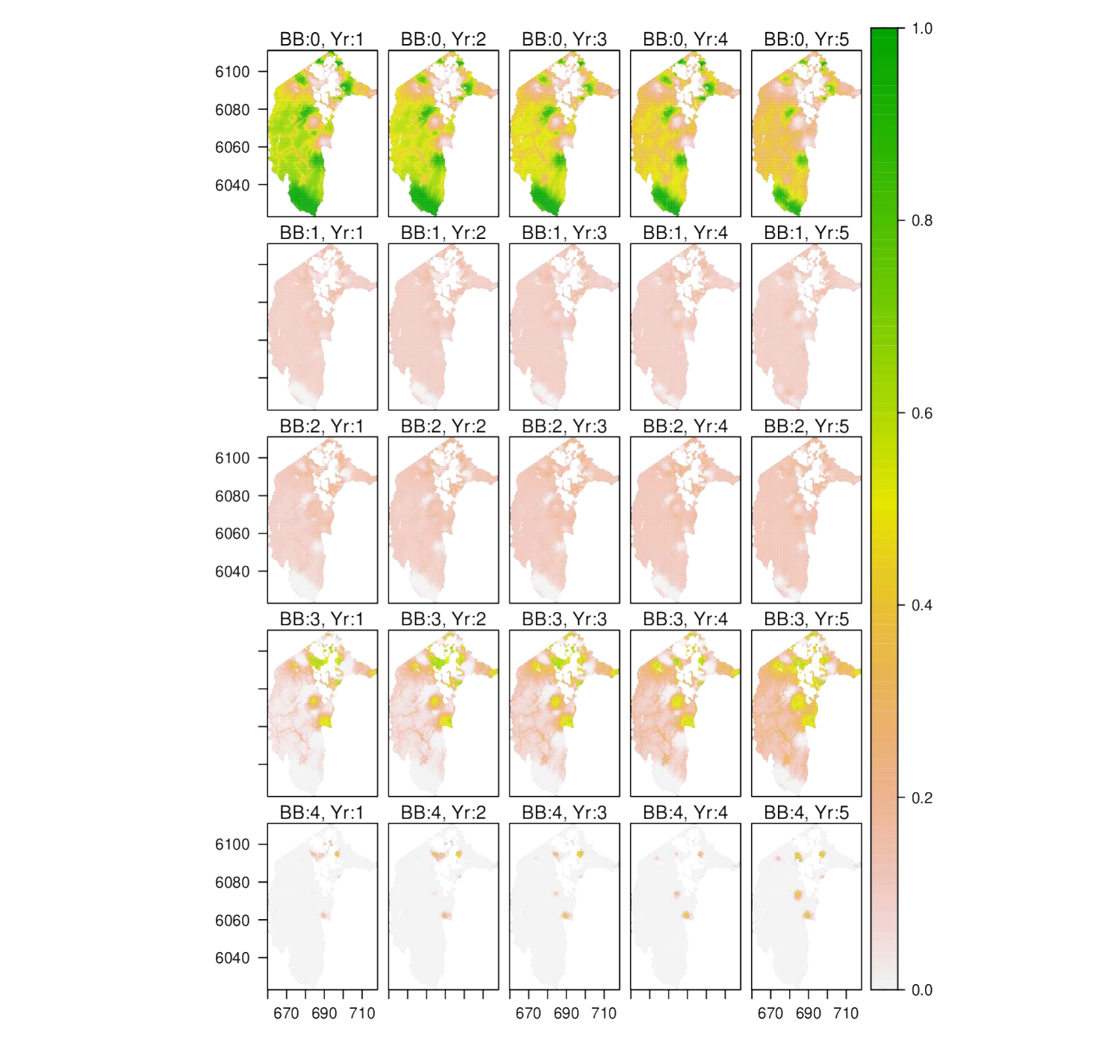}
	\caption{The predictive posterior median probability of the aggregated modified Braun-Blanquet scores for Saint John's wort by years (columns) and scores (rows). Based on data supplied by the ACT Government. \href{https://actmapi-actgov.opendata.arcgis.com/datasets/ACTGOV::actgov-border/about}{ACT border data} and \href{https://actmapi-actgov.opendata.arcgis.com/datasets/ACTGOV::actgov-vegetation-map-2018/about}{ACT vegetation data} used under \href{https://creativecommons.org/licenses/by/4.0/}{CC BY 4.0}.}\label{fig:StJW}
\end{figure}
	
\begin{figure}
	\includegraphics[width=0.95\textwidth]{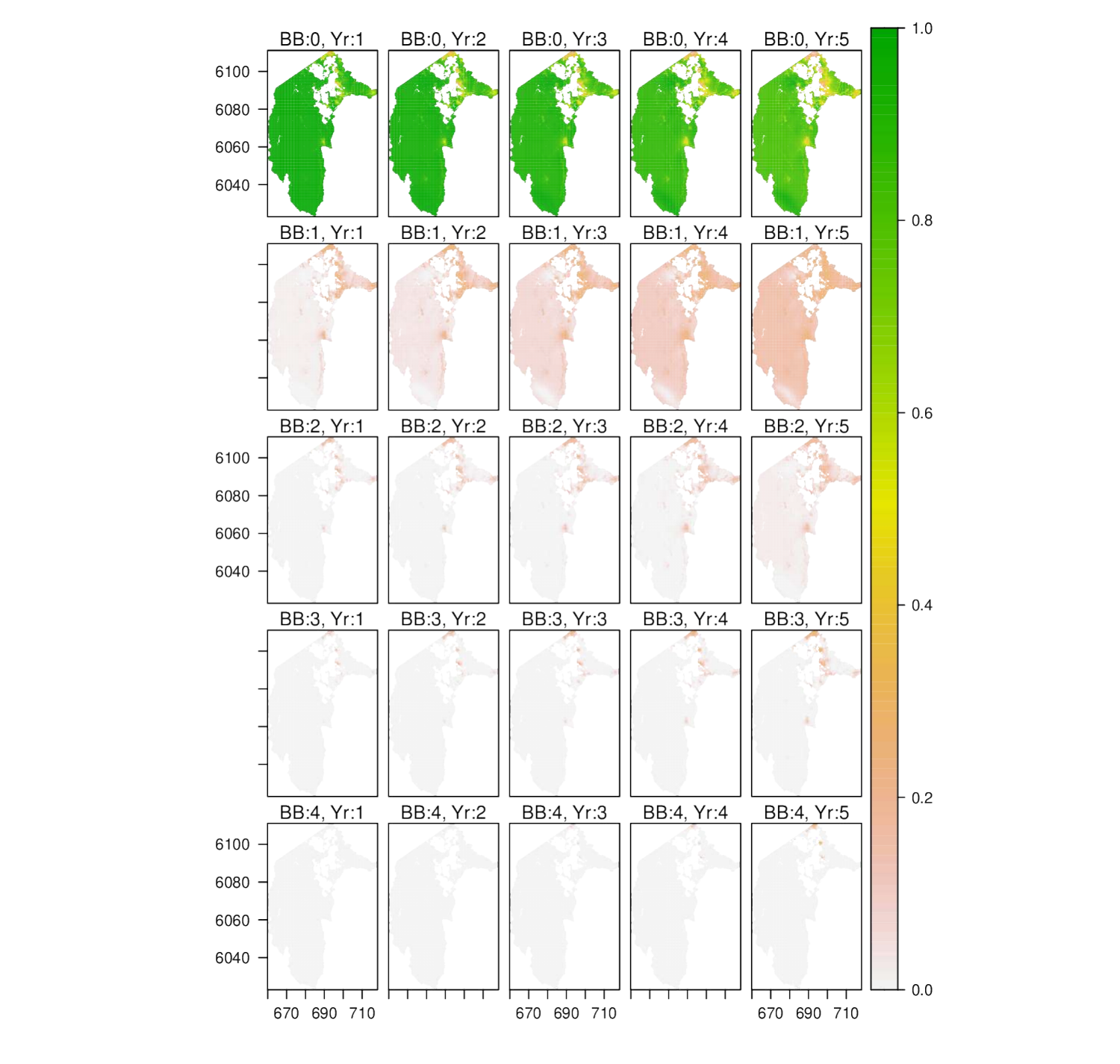}
	\caption{The predictive posterior median probability of the aggregated modified Braun-Blanquet scores for serrated tussock by years (columns) and scores (rows). Based on data supplied by the ACT Government. \href{https://actmapi-actgov.opendata.arcgis.com/datasets/ACTGOV::actgov-border/about}{ACT border data} and \href{https://actmapi-actgov.opendata.arcgis.com/datasets/ACTGOV::actgov-vegetation-map-2018/about}{ACT vegetation data} used under \href{https://creativecommons.org/licenses/by/4.0/}{CC BY 4.0}.}\label{fig:St}
\end{figure}

The sparseness of the ordinal data for each species in space and time (Section \ref{sec:data}) was accommodated by the separable spatio-temporal model formulation described above. The posterior estimates for parameters are summarised in Table \ref{tab:spp}. Maps of the predictive posterior median probability of each ordinal category on 1 July in each year for the four species are shown in Figures \ref{fig:AGC}, \ref{fig:CNG}, \ref{fig:StJW} and \ref{fig:St} (see Supplementary Information for estimated predictive posterior 0.05 and 0.95 quantiles). Annual predictions across the study area were based on the weed management control events as of 1 July in each year. Each species showed isolated patches for higher ordinal levels ranges that tended to increase in probability over the years. The estimated short effective spatial ranges may reflect the importance of fine-scale habitat feature or localised dispersal mechanisms for population growth. African lovegrass and Saint John's wort were negatively associated with forest habitat. A positive temporal trend was suggested for both African lovegrass and serrated tussock. A possible negative association with increasing distance from roads and trails  was suggested by negative posterior median estimates for both Chilean needle grass and St. John's wort, although the corresponding 95\% central credible intervals spanned zero.

The DSTSOM was motivated by the ecology of the weeds species (Section \ref{sec:Gompertz}) and the spatio-temporal aspect of targeted weeds management. However, the weeds observations that are sparse in space and time also show many absences. The low information content in the data contributes to DIC estimates (Table \ref{tab:dic}) that for some species indicate support for the simpler forms of Eq. \eqref{eq:eta_tau} without random effects ($M_1$) or the spatial only model ($M_2$, see Supplementary Information). The two species with lower relative DIC values for the DSTSOM ($M_3$) compared to the simpler models, St. John's wort and serrated tussock, also had the lowest median estimates for the temporal autocorrelation parameter. The high autocorrelations for all species indicate that the Gompertz growth model is approaching an exponential growth model (Section \ref{sec:Gompertz}) that would be expected for small populations far below carrying capacity where intraspecific density dependent interactions would be strongest. The weakest temporal autocorrelation parameter was estimated for St. John's wort that also had the greatest posterior median effective spatial range. Short effective spatial ranges relative to the study area (Figure \ref{fig:studyarea}) were estimated for all species.

The influence of spatio-temporal dependence was assessed on the posterior estimates of the coefficients for the weed control covariates, $\bm{\beta}_{C:p}$, including the association with species-specific control events and duration since a control event. For St. John's wort, the 95\% central CIs spanned zero for both duration ($\beta_{d\ StJW}$) and also the occurrence of a control event ($\beta_{ctrl\ StJW}$) in the simplest model $M_1$ without spatial or temporal dependence (Supplementary Information). The inclusion of spatial dependence led to a 95\% CI that was greater than zero for duration. Once temporal autocorrelation was incorporated then the presence of a control event was positively associated with coverage of St. John's wort but not duration since control (Table \ref{tab:spp}).  One conjectural  explanation for the latter positive association is that once population growth is accounted for in $M_3$, then a control event is more likely to occur with large population sizes that are more likely to be observed and hence targeted for control.  

Serrated tussock also suggested the possibility that estimated positive associations between control events and increased weed coverage  could arise due to higher probabilities of targeting high abundance sites. Posterior estimates (Table \ref{tab:spp}) suggested a positive association between population size and a control event that was followed by a return towards pre-treatment trend in population size (i.e., median estimates are positive for $\beta_{ctrl\ St}$ and negative for $\beta_{d\ St}$ in Table \ref{tab:spp}) in all three models of no dependence, spatial dependence and spatio-temporal dependence. For the two species with the strongest temporal autocorrelation, African lovegrass and Chilean needle grass, comparison of parameter estimates suggested that the inclusion of spatial random effects had the greatest effect on posterior median estimates for the coefficients, whereas the additional inclusion of temporal random effects had a lesser impact. The parameters for the impacts of control event occurrence on both species had 95\% central credible intervals (CIs) greater than zero in the GLM model without random effects, but these 95\% CIs included zero once accounting for either spatial dependence (Supplementary Information) or spatio-temporal dependence (Table \ref{tab:spp}). 

\section{Discussion}\label{sec:discussion}

Bayesian  models for ordinal data require sophisticated computational machinery to facilitate estimation in high-dimensional contexts with non-Gaussian likelihoods \citep{Schliep2015, Cameletti2017}. This study demonstrates inference of high-dimensional spatio-temporal models for ordinal data, and is applicable to both longitudinal time series models and spatial models  as special cases. The sequential ordinal model is shown to have pragmatic properties that permit estimation and prediction within existing spatio-temporal model frameworks developed for univariate data (Section \ref{sec:DSTSOM}), thereby forestalling the need to account for the additional complexity of multivariate data likelihood structures. The dynamic sequential ordinal model is shown to be a particularly apt choice for biological populations (Section \ref{sec:Gompertz}). Desirable properties such as collapsibility with respect to ordinal categories can be accommodated (Section \ref{sec:prophaz}). Estimation and prediction for spatio-temporal models proceeds as for univariate data analysis, and is easily accessible for example by the popular INLA software package (Section \ref{sec:INLA}). The application of a separable dynamic spatio-temporal ordinal model (DSTSOM) to invasive weeds ordinal data illustrated the importance of taking into account spatio-temporal dependence while estimating statistical relationships with weed control effort (Section \ref{sec:application}). 

The proposed approach recasts the multivariate ordinal model into a univariate estimation problem, and thus allows the application of standard univariate spatio-temporal modelling procedures as demonstrated with a separable autoregressive space-time model for the latent random effects \citep{Cressie2011, Cameletti2013, Blangiardo2015}. Extensions to non-separable latent space-time process models could be implemented by including more complex structure in the state transition matrix of Eq. \eqref{eq:eta_tau} \citep[see][for examples]{Cressie2011}. Such extensions are facilitated by the removal of constraints on the threshold parameters through the sequential ordinal model that may also improve MCMC-based estimation methods. MCMC methods otherwise require special attention for the constrained ordering of the cut or threshold parameters in spatial and spatio-temporal Bayesian cumulative ordinal models \citep{Schliep2015, Cameletti2017, Erhardt2024}. 

Bayesian prior elicitation also benefits from the  unconstrained threshold parameters in the sequential ordinal model that enable multivariate normal or Student's t prior elicitation for unknown coefficients in ordinal GLMs \citep{Hosack2024}. The inclusion of informative priors into Bayesian analysis is an important challenge for INLA \citep{Rue2017}, and statistical models such as GLMs more generally \citep{Mikkola2024}. Incorporation of ecological domain knowledge into the priors of Bayesian dynamic spatio-temporal sequential ordinal model through expert elicitation further requires treatment of the random effects, perhaps by incorporating random effects using a multivariate t distribution for marginal predictions of the sequential probabilities. Prior elicitation for the Bayesian DSTSOM would therefore be an interesting avenue of future research. 

Extensions are also possible to multivariate models that incorporate other response data types, such as continuous and Poisson count data, into a joint likelihood structure. The complementary log-log link function used in the sequential ordinal model application here (Section \ref{sec:prophaz}), for example, also maps into Poisson point process models with binary likelihoods \citep{Baddeley2010}. The Poisson point process features widely in both spatial and dynamic spatio-temporal species distribution models with joint likelihoods \citep{Fletcher2019} for which INLA has proven useful  \citep{Martinez-Minaya2018, Isaac2020, Foster2024}.  Spatio-temporal models with shared or correlated random effects are also of interest, such as generalised spatial dynamic factor models \citep{Gamerman2013}, as might arise in a multispecies setting with strong spatio-temporal dependence. INLA has been demonstrated for multivariate binomial responses in such models, for example \citep{Kifle2017}. The binary likelihood representation of the dynamic spatio-temporal sequential ordinal model therefore also allows application of these multivariate techniques to joint models of ordinal datasets.

\section*{Acknowledgements}

The authors thanks the many additional ACT Government staff and contractors that have contributed to collecting the field data used in this study. The authors thank Scott Foster and David Peel for comments and feedback.

\section*{Supplementary Information}

Supplementary information is provided in the vignette of the \textbf{DSTSOM} \textsf{R} package \citep{DSTSOM2025} available at \url{https://github.com/csiro/DSTSOM}.

\bibliographystyle{apalike}
\bibliography{ordinal_analysis}

\begin{thebibliography}{}

\bibitem[ACT, 2016]{ACT2016}
ACT (2016).
\newblock {\em ACT Biosecurity Strategy 2016--2026}.
\newblock Australian Capital Territory Government.
\newblock \url{https://www.act.gov.au/open/act-biosecurity-strategy}.

\bibitem[Ansong and Pickering, 2013]{Ansong2013}
Ansong, M. and Pickering, C. (2013).
\newblock Are weeds hitchhiking a ride on your car? {A} systematic review of
  seed dispersal on cars.
\newblock {\em PLoS ONE}, 8:e80275.

\bibitem[Armstrong and Sloan, 1989]{Armstrong1989}
Armstrong, B.~G. and Sloan, M. (1989).
\newblock Ordinal regression models for epidemiologic data.
\newblock {\em American Journal of Epidemiology}, 129:191--204.

\bibitem[Baddeley et~al., 2010]{Baddeley2010}
Baddeley, A., Berman, M., Fisher, N.~I., Hardegen, A., Milne, R.~K.,
  Schuhmacher, D., Shah, R., and Turner, R. (2010).
\newblock Spatial logistic regression and change-of-support in poisson point
  processes.
\newblock {\em Electronic Journal of Statistics}, 4:1151--1201.

\bibitem[Bahn et~al., 2008]{Bahn2008}
Bahn, V., Krohn, W.~B., and O’Connor, R.~J. (2008).
\newblock Dispersal leads to spatial autocorrelation in species distributions:
  A simulation model.
\newblock {\em Ecological Modelling}, 213:285--292.

\bibitem[Baker, 1994]{Baker1994}
Baker, S.~G. (1994).
\newblock The multinomial--{P}oisson transformation.
\newblock {\em The Statistician}, 43:495--504.

\bibitem[Bender and Benner, 2000]{Bender2000}
Bender, R. and Benner, A. (2000).
\newblock Calculating ordinal regression models in {SAS} and {S-Plus}.
\newblock {\em Biometrical Journal}, 42:677--699.

\bibitem[Berridge and Whitehead, 1991]{Berridge1991}
Berridge, D.~M. and Whitehead, J. (1991).
\newblock Analysis of failure time data with ordinal categories of response.
\newblock {\em Statistics in Medicine}, 10:1703--1710.

\bibitem[Blangiardo and Cameletti, 2015]{Blangiardo2015}
Blangiardo, M. and Cameletti, M. (2015).
\newblock {\em Spatial and Spatio-Temporal Bayesian Models with R-INLA}.
\newblock John Wiley \& Sons, Ltd., Chichester, UK.

\bibitem[Braun-Blanquet, 1932]{Braun-Blanquet1932}
Braun-Blanquet, J. (1932).
\newblock {\em Plant sociology. The Study of Plant Communities}.
\newblock McGraw Hill, New York.
\newblock Translated, revised and edited by George D. Fuller and Henry S.
  Conard.

\bibitem[Bruelheide et~al., 2019]{Bruelheide2019}
Bruelheide, H., Dengler, J., Jim\'{e}nez-Alfaro, B., Purschke, O., Hennekens,
  S.~M., Chytr\'{y}, M., Pillar, V.~D., Jansen, F., Kattge, J., et~al. (2019).
\newblock {sPlot} – a new tool for global vegetation analyses.
\newblock {\em Journal of Vegetation Science}, 30:161--186.

\bibitem[Cameletti et~al., 2017]{Cameletti2017}
Cameletti, M., De~Rubeis, V., Ferrari, C., Sbarra, P., and Tosi, P. (2017).
\newblock An ordered probit model for seismic intensity data.
\newblock {\em Stochastic Environmental Research and Risk Assessment},
  31:1593--1602.

\bibitem[Cameletti et~al., 2013]{Cameletti2013}
Cameletti, M., Lindgren, F., Simpson, D., and Rue, H. (2013).
\newblock Spatio-temporal modeling of particulate matterconcentration through
  the {SPDE} approach.
\newblock {\em Advances in Statistical Analysis}, 97:109--131.

\bibitem[Christen and Matlack, 2006]{Christen2006}
Christen, D. and Matlack, G. (2006).
\newblock The role of roadsides in plant invasions: {A} demographic approach.
\newblock {\em Conservation Biology}, 20:385--391.

\bibitem[Cressie and Wikle, 2011]{Cressie2011}
Cressie, N. and Wikle, C.~K. (2011).
\newblock {\em Statistics for Spatio-Temporal Data}.
\newblock John Wiley and Sons, Inc., Hoboken, New Jersey.

\bibitem[CSIS, 2024a]{CSIS2024}
CSIS (2024a).
\newblock Profile {African Lovegrass}.
\newblock Centre for Invasive Species Solutions.
  \url{https://weeds.org.au/profiles/african-lovegrass-weeping/} [Accessed: 13
  June 2025].

\bibitem[CSIS, 2024b]{CSISCNG2024}
CSIS (2024b).
\newblock Profile chilean needle grass.
\newblock Centre for Invasive Species Solutions.
  \url{https://weeds.org.au/profiles/chilean-needle-grass/} [Accessed: 9 July
  2025].

\bibitem[CSIS, 2024c]{CSIS2024SJW}
CSIS (2024c).
\newblock Profile {St. John's Wort}.
\newblock Centre for Invasive Species Solutions.
  \url{https://weeds.org.au/profiles/st-johns-wort/} [Accessed: 13 June 2025].

\bibitem[CSIS, 2024d]{CSIS2024Serr}
CSIS (2024d).
\newblock Serrated tussock, yass river tussock, yass tussock, nassella tussock
  (nz).
\newblock Centre for Invasive Species Solutions.
  \url{https://weeds.org.au/profiles/serrated-tussock-yass/} [Accessed: 9 July
  2025].

\bibitem[Dambly et~al., 2023]{Dambly2023}
Dambly, L.~I., Isaac, N.~J., Jones, K.~E., Boughey, K.~L., and O'Hara, R.~B.
  (2023).
\newblock Integrated species distribution models fitted in {INLA} are sensitive
  to mesh parameterisation.
\newblock {\em Ecography}, 2023:e06391.

\bibitem[Dennis et~al., 2006]{Dennis2006}
Dennis, B., Ponciano, J.~M., R., S., Taper, M.~L., and Staples, D.~F. (2006).
\newblock Estimating density dependence, process noise, and observation error.
\newblock {\em Ecological Monographs}, 76:323--341.

\bibitem[Erhardt et~al., 2024]{Erhardt2024}
Erhardt, R., Hepler, S., Wolodkin, D., and Greene, A. (2024).
\newblock Spatio-temporal forecasting for the {US} drought monitor.
\newblock {\em Journal of the Royal Statistical Society Series C: Applied
  Statistics}, 73:1203--1220.

\bibitem[Fahrmeir and Tutz, 1994]{Fahrmeir1994}
Fahrmeir, L. and Tutz, G. (1994).
\newblock {\em Multivariate Statistical Modelling Based on Generalized Linear
  Models}.
\newblock Springer-Verlag New York Inc.

\bibitem[Fletcher et~al., 2019]{Fletcher2019}
Fletcher, Jr., R.~J., Hefley, T.~J., Robertson, E.~P., Zuckerberg, B.,
  McCleery, R.~A., and Dorazio, R.~M. (2019).
\newblock A practical guide for combining data to model species distributions.
\newblock {\em Ecology}, 100:e02710.

\bibitem[Foster et~al., 2024]{Foster2024}
Foster, S.~D., Peel, D., Hosack, G.~R., Hoskins, A., Mitchell, D.~J., Proft,
  K., Yang, W.-H., Uribe-Rivera, D.~E., and Froese, J.~G. (2024).
\newblock ‘{RISDM} ‘: {S}pecies distribution modelling from multiple data
  sources in {R}.
\newblock {\em Ecography}, 6:e06964.

\bibitem[Fuglstad et~al., 2019]{Fuglstad2019}
Fuglstad, G., Simpson, D., Lindgren, F., and Rue, H. (2019).
\newblock Constructing priors that penalize the complexity of {G}aussian random
  fields.
\newblock {\em Journal of the American Statistical Association}, 114:445--452.

\bibitem[Gamerman and Salazar, 2013]{Gamerman2013}
Gamerman, D. and Salazar, E. (2013).
\newblock Hierarchical modelling in time series: the factor analytic approach.
\newblock In {\em Bayesian Theory and Applications}, pages 167--182. Oxford
  University Press, Oxford.

\bibitem[Gelfand et~al., 2003]{Gelfand2003}
Gelfand, A.~E., Kim, H.-J., Sirmans, C.~F., and Banerjee, S. (2003).
\newblock Spatial modeling with spatially varying coefficient processes.
\newblock {\em Journal of the American Statistical Association}, 98:387--396.

\bibitem[G{\"o}b et~al., 2007]{Gob2007}
G{\"o}b, R., McCollin, C., and Ramalhoto, M.~F. (2007).
\newblock Ordinal methodology in the analysis of {L}ikert scales.
\newblock {\em Quality and Quantity}, 41:601--626.

\bibitem[G{\'o}mez-Rubio, 2020]{Gomez-Rubio2020}
G{\'o}mez-Rubio, V. (2020).
\newblock {\em Bayesian Inference with INLA}.
\newblock CRC Press, Boca Raton, FL.

\bibitem[Guisan and Harrell, 2000]{Guisan2000}
Guisan, A. and Harrell, F.~E. (2000).
\newblock Ordinal response regression models in ecology.
\newblock {\em Journal of Vegetation Science}, 11:617--626.

\bibitem[Harvey, 1989]{Harvey1989}
Harvey, A.~C. (1989).
\newblock {\em Forecasting, Structural Time Series Models and the Kalman
  Filter}.
\newblock Cambridge University Press, Cambridge, UK.

\bibitem[Higgs and Hoeting, 2010]{Higgs2010}
Higgs, M.~D. and Hoeting, J.~A. (2010).
\newblock A clipped latent variable model for spatially correlated ordered
  categorical data.
\newblock {\em Computational Statistics and Data Analysis}, 54:1999--2011.

\bibitem[Hosack, 2024]{Hosack2024}
Hosack, G.~R. (2024).
\newblock Prior elicitation for generalised linear models and extensions.
\newblock {\em Bayesian Analysis}.
\newblock Advance Publication 1 - 30. \url{https://doi.org/10.1214/24-BA1472}.

\bibitem[Hosack et~al., 2025]{DSTSOM2025}
Hosack, G.~R., Yang, W.-H., and Pike, K.~N. (2025).
\newblock {\em DSTSOM: Dynamic Spatio-Temporal Sequential Ordinal Models}.
\newblock R package version 0.1.0. \url{https://github.com/csiro/DSTSOM}.

\bibitem[Ip and Wu, 2024]{Ip2024}
Ip, R. H.~L. and Wu, K. Y.~K. (2024).
\newblock A {M}arkov random field model with cumulative logistic functions for
  spatially dependent ordinal data.
\newblock {\em Journal of Applied Statistics}, 51:70--86.

\bibitem[Isaac et~al., 2020]{Isaac2020}
Isaac, N.~J., Jarzyna, M.~A., Keil, P., Dambly, L.~I., Boersch-Supan, P.~H.,
  Browning, E., Freeman, S.~N., Golding, N., Guillera-Arroita, G., Henrys,
  P.~A., Jarvis, S., Lahoz-Monfort, J., Pagel, J., Pescott, O.~L., Schmucki,
  R., Simmonds, E.~G., and O’Hara, R.~B. (2020).
\newblock Data integration for large-scale models of species distributions.
\newblock {\em Trends in Ecology and Evolution}, 35:56--67.

\bibitem[Ivanova, 2024]{Ivanova2024}
Ivanova, N. (2024).
\newblock Global overview of the application of the {Braun-Blanquet} approach
  in research.
\newblock {\em Forests}, 15:937.

\bibitem[Ives et~al., 2003]{Ives2003}
Ives, A.~R., Dennis, B., Cottingham, K.~L., and Carpenter, S.~R. (2003).
\newblock Estimating community stability and ecological interactions from
  time-series data.
\newblock {\em Ecological Monographs}, 73:301--330.

\bibitem[Jung and Lee, 2012]{Jung2012}
Jung, I. and Lee, H. (2012).
\newblock Spatial cluster detection for ordinal outcome data.
\newblock {\em Statistics in Medicine}, 31:4040--4048.

\bibitem[Kifle et~al., 2017]{Kifle2017}
Kifle, Y.~W., Hens, N., and Faes, C. (2017).
\newblock Cross-covariance functions for additive and coupled joint
  spatiotemporal {SPDE} models in {R-INLA}.
\newblock {\em Environmental and Ecological Statistics}, 24:551--586.

\bibitem[Krainski et~al., 2019]{Krainski2019}
Krainski, E., G{\'o}mez-Rubio, V., Bakka, H., Lenzi, A., Castro-Camilo, D.,
  Simpson, D., Lindgren, F., and Rue, H. (2019).
\newblock {\em Advanced Spatial Modeling with Stochastic Partial Differential
  Equations Using R and INLA}.
\newblock CRC Press, Boca Raton, FL.

\bibitem[Laara and Matthews, 1985]{Laara1985}
Laara, E. and Matthews, J. N.~S. (1985).
\newblock The equivalence of two models for ordinal data.
\newblock {\em Biometrika}, 72:206--207.

\bibitem[Lindgren and Rue, 2015]{Lindgren2015}
Lindgren, F. and Rue, H. (2015).
\newblock Bayesian spatial modelling with {R-INLA}.
\newblock {\em Journal of Statistical Software}, 63(19):1--25.

\bibitem[Linero et~al., 2018]{Linero2018}
Linero, A.~R., Bradley, J.~R., and Desai, A. (2018).
\newblock Multi-rubric models for ordinal spatial data with application to
  online ratings data.
\newblock {\em Annals of Applied Statistics}, 12:2054--2074.

\bibitem[Liu et~al., 2019]{Liu2019}
Liu, J., Hainen, A., Li, X., Nie, Q., and Nambisan, S. (2019).
\newblock Pedestrian injury severity in motor vehicle crashes: An integrated
  spatio-temporal modeling approach.
\newblock {\em Accident Analysis and Prevention}, 132:105272.

\bibitem[Mart{\'i}nez-Minaya et~al., 2018]{Martinez-Minaya2018}
Mart{\'i}nez-Minaya, J., Cameletti, M., Conesa, D., and Pennino, M.~G. (2018).
\newblock Species distribution modeling: a statistical review with focus in
  spatio-temporal issues.
\newblock {\em Stochastic Environmental Research and Risk Assessment},
  32:3227--3244.

\bibitem[Mart{\'i}nez-Minaya and Rue, 2024]{Martinez-Minaya2024}
Mart{\'i}nez-Minaya, J. and Rue, H. (2024).
\newblock A flexible {B}ayesian tool for {CoDa} mixed models: logistic-normal
  distribution with {D}irichlet covariance.
\newblock {\em Statistics and Computing}, 34:116.

\bibitem[Martins et~al., 2013]{Martins2013}
Martins, T.~G., Simpson, D., Lindgren, F., and Rue, H. (2013).
\newblock Bayesian computing with {INLA}: New features.
\newblock {\em Computational Statistics and Data Analysis}, 67:68--83.

\bibitem[McCullagh, 1980]{McCullagh1980}
McCullagh, P. (1980).
\newblock Regression models for ordinal data.
\newblock {\em Journal of the Royal Statistical Society. Series B
  (Methodological)}, 42:109--142.

\bibitem[McCullagh and Nelder, 1989]{McCullagh1989}
McCullagh, P. and Nelder, J.~A. (1989).
\newblock {\em {Generalized Linear Models}}.
\newblock Chapman \& Hall/CRC, Boca Raton, Florida USA, 2nd edition.

\bibitem[Mikkola et~al., 2024]{Mikkola2024}
Mikkola, P., Martin, O.~A., Chandramouli, S., Hartmann, M., Pla, O.~A., Thomas,
  O., Pesonen, H., Corander, J., Vehtari, A., Kaski, S., B\"{u}rkner, P.-C.,
  and Klami, A. (2024).
\newblock Prior knowledge elicitation: {T}he past, present, and future.
\newblock {\em Bayesian Analysis}, 19:1129 --1161.

\bibitem[Podani, 2006]{Podani2006}
Podani, J. (2006).
\newblock {Braun-Blanquet}'s legacy and data analysis in vegetation.
\newblock {\em Journal of Vegetation Science}, 17:113--117.

\bibitem[Righetto et~al., 2020]{Righetto2020}
Righetto, A.~J., Faes, C., Vandendijck, Y., and Ribeiro, Jr., P.~J. (2020).
\newblock On the choice of the mesh for the analysis of geostatistical data
  using {R-INLA}.
\newblock {\em Communications in Statistics -- Theory and Methods},
  49:203--220.

\bibitem[Rue et~al., 2009]{Rue2009}
Rue, H., Martino, S., and Chopin, N. (2009).
\newblock Approximate {B}ayesian inference for latent {G}aussian models by
  using integrated nested {L}aplace approximations.
\newblock {\em Journal of the Royal Statistical Society Series B}, 71:319--392.

\bibitem[Rue et~al., 2017]{Rue2017}
Rue, H., Riebler, A., S\o{}rbye, S.~H., Illian, J.~B., Simpson, D.~P., and
  Lindgren, F.~K. (2017).
\newblock Bayesian computing with {INLA}: A review.
\newblock {\em Annual Review of Statistics and Its Application}, 4:395--421.

\bibitem[SACTCG, 2023]{SAGC2023}
SACTCG (2023).
\newblock {\em ACT Weeds Manual}.
\newblock Southern ACT Catchment Group.
\newblock \url{https://sactcg.org.au/act-weeds-manual/} [Accessed: 13 June
  2025].

\bibitem[Schliep and Hoeting, 2015]{Schliep2015}
Schliep, E.~M. and Hoeting, J.~A. (2015).
\newblock Data augmentation and parameter expansion for independent or
  spatially correlated ordinal data.
\newblock {\em Computational Statistics and Data Analysis}, 90:1--14.

\bibitem[Spiegelhalter et~al., 2002]{Spiegelhalter2002}
Spiegelhalter, D.~J., Best, N.~G., Carlin, B.~P., and Van Der~Linde, A. (2002).
\newblock Bayesian measures of model complexity and fit.
\newblock {\em Journal of the Royal Statistical Society: Series B (Statistical
  Methodology)}, 64:583--639.

\bibitem[Thorson et~al., 2015]{Thorson2015}
Thorson, J.~T., Skaug, H.~J., Kristensen, K., Shelton, A.~O., Ward, E.~J.,
  Harms, J.~H., and Benante, J.~A. (2015).
\newblock The importance of spatial models for estimating the strength of
  density dependence.
\newblock {\em Ecology}, 96:1202--1212.

\bibitem[Tutz, 1991]{Tutz1991}
Tutz, G. (1991).
\newblock Sequential models in categorical regression.
\newblock {\em Computational Statistics and Data Analysis}, 11:275--295.

\bibitem[Tutz, 2012]{Tutz2012}
Tutz, G. (2012).
\newblock {\em Regression for Categorical Data}.
\newblock Cambridge University Press, New York.

\bibitem[Verdoy, 2021]{Verdoy2021}
Verdoy, P.~J. (2021).
\newblock Enhancing the {SPDE} modeling of spatial point processes with {INLA},
  applied to wildfires. choosing the best mesh for each database.
\newblock {\em Communications in Statistics-Simulation and Computation},
  50(10):2990--3030.

\bibitem[West and Harrison, 1997]{West1997}
West, M. and Harrison, J. (1997).
\newblock {\em Bayesian Forecasting and Dynamic Models}.
\newblock Springer, New York, 2nd edition.

\bibitem[Zeil-Rolfe et~al., 2024]{Zeil-Rolfe2024}
Zeil-Rolfe, I., O'Loughlin, L., and Gooden, B. (2024).
\newblock Habitat context and functional growth traits explain alien plant
  invader impacts on native vegetation communities.
\newblock {\em Biological Invasions}, 26:2663--2679.

\end{thebibliography}

\end{document}